\documentclass[10pt, letter, showpacs, titlepage, showkeys, floatfix, aps]{revtex4-1}
\usepackage{times}
\usepackage{hyperref}
\usepackage{amsmath}
\usepackage{amssymb}
\usepackage{graphicx}
\usepackage{xcolor}
\usepackage{booktabs}
\usepackage{rotating}
\usepackage{bm}
\usepackage{float}
\usepackage{epstopdf}
\usepackage{comment}
\makeatletter

\newcommand{\Rmnum}[1]{\expandafter\@slowromancap\romannumeral #1@}
\makeatother

\usepackage{dirtytalk} % For some quotations and other gramatical representations.
\usepackage{caption}
\usepackage{subcaption}

\begin{document}
\title{RFOX (Rotated-Field Oscillatory eXchange) quantum algorithm: Towards Parameter-Free Quantum Optimizers}
%\author{Brian García Sarmina, Guo-Hua Sun and Shi-Hai Dong}
\author{Brian García Sarmina$^{1,2}$}
\email[E-mail:]{brian.garsar.6@gmail.com}

\author{Guo-Hua Sun$^{1}$}
\email[E-mail:]{sunghdb@yahoo.com}

\author{Shi-Hai Dong$^{1,3}$}
\email[E-mail:]{dongsh2@yahoo.com}

\affiliation{$^1$ Centro de Investigaci\'{o}n en Computaci\'{o}n, Instituto Polit\'{e}cnico Nacional, UPALM, CDMX 07738, México.}
\affiliation{$^2$ Centro de Tecnolog\'{i}as en C\'{o}mputo y Comunicaci\'{o}n, FESC - Universidad Nacional Aut\'{o}noma de M\'{e}xico, Cuautitl\'{a}n Izcalli, 54714, M\'{e}xico.}    
\affiliation{$^3$ Research Center for Quantum Physics, Huzhou University, Huzhou 313000, China.}

\begin{abstract}
We introduce RFOX (Rotated-Field Oscillatory eXchange), a parameter-free quantum algorithm for combinatorial optimization that combines an almost constant non-stoquastic $XX$ catalyst with a weak harmonic $ZX$ counter-diabatic term. Using the Floquet-Magnus expansion, we derive an effective Hamiltonian whose leading-order $\mathcal{O}(\delta/\omega)$ corrections yield local $Y$ fields, field-modulated 2-body terms, and poly-local 3-body topological interactions driven by graph connectivity. This structure ensures a nearly flat instantaneous spectral gap, preventing the unpredictable gap collapses typical of conventional $X$ (stoquastic), $XX$, and $X+sXX$ (non-stoquastic) driver schedules. Extensive noiseless simulations and physical hardware experiments on IBM Quantum processors (up to 20 qubits) validate our spectral predictions. RFOX consistently attains near-optimal or exact ground states in the random-field Ising model using up to an order of magnitude fewer Trotter slices, with an advantage that grows alongside problem disorder. These results suggest that fixed-gap, non-stoquastic drivers augmented with analytically derived counter-diabatic terms offer a scalable, tuning-free route for quantum optimization.
\end{abstract}

\maketitle

\section{Introduction}

Inside the quantum computing paradigm, the Variational Quantum Algorithms (VQAs) have become a cornerstone approach for harnessing near-term quantum devices in the noisy intermediate-scale quantum (NISQ) era. Landmark schemes such as the Variational Quantum Eigensolver (VQE) \cite{peruzzo2014variational, fedorov2022vqe, liu2022layer} and the Quantum Approximate Optimization Algorithm (QAOA) \cite{farhi2014quantum, blekos2024review} exemplify how shallow, parameterized circuits, optimized through a quantum-classical feedback loop, can already tackle demanding problems in quantum chemistry and combinatorial optimization that strain classical resources \cite{zhou2020quantum, zhou2023qaoa, crooks2018performance, bae2024recursive}. Beyond these initial demonstrations, a rapidly expanding body of work continues to expand the reach of VQAs, driving progress in ansatz engineering, error-resilient circuit design, and strategies for effective parameter initialization \cite{Preskill2018NISQ, Peruzzo2014VQE, Cerezo2021Review, Motta2019QITE}.

Despite these successes, the hybrid nature of VQAs imposes fundamental bottlenecks. As the system size increases, the classical optimizer encounters barren plateaus, regions where cost-function gradients vanish exponentially, which delay training even in noiseless simulations \cite{grover1996fast, schuld2014quantum, haug2021optimal, zhang2022escaping}. In real hardware, stochastic gate errors further flatten the optimization landscape and introduce shot noise, overwhelming gradient signals, and making large-scale optimization prohibitively expensive in both circuit executions and classical processing time \cite{McClean2018Barren, Wang2021NoiseBarren}.

To address these limitations, we introduce \emph{Rotated-Field Oscillatory eXchange} (RFOX) quantum algorithm, an entirely quantum protocol tailored to the random-field Ising model (RFIM) problem. Instead of alternating between quantum circuits and a classical optimizer, RFOX directly encodes the local fields via phase rotations and unifies the cost and mixing dynamics within a single two‐qubit gate sequence, $U(\theta, \phi) = RZX(\theta) \ RXX(\phi)$, applied to each edge of the RFIM graph. By setting $\theta,\phi \propto \Delta t$, this construction implements a first‐order Trotter step inspired by the quantum adiabatic theorem. To enhance adiabaticity, both gates are modulated by a small harmonic amplitude $\delta \ll 1$, synchronized with $\Delta t$, generating counter‐diabatic corrections while preserving an almost constant non-stoquastic gap.

In disordered systems such as the RFIM, where frustrated couplings and random longitudinal fields hinder conventional quantum annealing, introducing a transverse interaction channel $X \otimes X$ (XX) can accelerate convergence and suppress localization-induced gap closures, outperforming pure transverse field schedules \cite{Suzuki2007RFIM, Zhou2022RFIM, Yang2024QAOAParams}. A key design element of RFOX is the combination of a nearly constant non‐stoquastic XX driver with a counter-diabatic term of time oscillation $Z \otimes X$ (ZX). The XX interaction alleviates the stoquastic bottleneck by coupling computational basis states differing by multiple spin flips, thus widening the spectral gap \cite{lykiardopoulou2021improving, albash2019role, choi2021essentiality}. However, a large gap alone does not guarantee adiabatic evolution: when problem fields vary rapidly, residual diabatic transitions can accumulate. 

The harmonic ZX term serves as the driver for the leading-order component of the adiabatic gauge potential, modulated by a phase-shifted sinusoidal envelope. As derived via the Floquet-Magnus expansion, this supplies precisely the time derivative required to cancel non-adiabatic couplings. Crucially, these leading-order corrections scale as $\mathcal{O}(\delta/\omega)$ and manifest as a richer physical structure than previously assumed: they generate local spin rotations, field-modulated 2-body terms, and poly-local 3-body topological interactions dictated by the graph connectivity, without adding any static cost term that could close the gap. As a result, the combined XX+ZX driver preserves a nearly constant minimum gap while suppressing diabatic leakage, enabling the system to reach approximate, and in some cases exact, ground states at fixed, shallow depth and without classically tuned parameters.

Simulations in 7-12 spin RFIM instances show that a circuit depth of $p=100$ consistently yields near-optimal, and often exact, ground states, outperforming all other tested methods. We further examine the convergence and scalability of RFOX in randomly generated RFIM instances with 12 to 20 qubits executed on IBM Quantum backends: \texttt{ibm\_brisbane}, \texttt{ibm\_sherbrooke}, and \texttt{ibm\_torino}. Compared with conventional drivers, either an XX-only catalyst or an X + sXX schedule, RFOX achieves higher ground-state fidelities and faster convergence. By eliminating the classical optimization loop, the protocol reduces operational overhead, improving both the practicality and scalability of quantum optimization on near-term hardware.

Section II reviews the theoretical background of the algorithm. Section III provides a formal description of RFOX. Section IV details the generation of RFIM instances and the implementation of these problems on quantum hardware. Section V presents our simulation and experimental results, and Section VI concludes the paper.

\section{Theoretical Background}

In this section, we outline the main concepts underlying the RFOX quantum algorithm, which will be detailed in the following section.

RFIM problems can be represented as an undirected graph $G = (V, E)$ with $|V| = N$ sites (or as a $d$-dimensional lattice when $G$ is regular), where each classical Ising spin $\sigma_i \in \{-1, +1\}$ for $i \in V$. The RFIM Hamiltonian \cite{cipra1987introduction, hughes1999ising} is:
\begin{equation}
    H = - \sum_{\langle i, j \rangle \in E}^{} J_{ij}\sigma_{i} \sigma_{j} - \sum_{i \in V}^{} h_{i}\sigma_{i}
\end{equation}

Here, $J_{ij}$ denotes the coupling strength, positive for ferromagnetic interactions ($J_{ij}>0$) and negative for antiferromagnetic interactions ($J_{ij}<0$), $h_i$ are independent magnetic fields, and $\langle i, j\rangle$ indicates nearest-neighbor pairs.

This Hamiltonian has two energy contributions:
\begin{itemize}
  \item The \emph{interaction energy} $-\sum_{\langle i, j\rangle \in E} J_{ij} \sigma_i \sigma_j$, favors aligning spins to minimize the pairwise coupling energy.
  \item The \emph{field energy} $-\sum_{i \in V} h_i \sigma_i$, favors each spin aligning with its local random field.
\end{itemize}

\subsection{Magnetic field encoding using phase gates}

In RFOX, we develop a \emph{phase mapping} of the magnetic field energy. The term  
$\sum_{j\in V} h_{j}\,\sigma_{j}$ is mapped to phase rotations $P(\phi_{j})$ via a sequence of phase gates that create an interference pattern between two Walsh-Hadamard transforms:
\begin{equation}
 H_{B} = H^{\otimes n} P(\phi_{j}) H^{\otimes n} = e^{-i\frac{\phi_{j}}{2} H Z H} \ .
\end{equation}
Here, $H^{\otimes n}$ denotes the Walsh-Hadamard transform over $n$ qubits. See Appendix A for details.

This phase encoding generates an initial candidate ground state (or states) by quantum interference, accounting only for the field energy contribution.

% This phase encoding produces an initial potential ground state (or states) by quantum interference, taking into account only the field energy contribution.

\subsection{Stoquastic vs. non-stoquastic Hamiltonians}

To encode the interaction energy term from RFIM in RFOX, we leverage principles from the adiabatic theorem, quantum walks, non-stoquastic Hamiltonians, and counter-diabatic driving.

A stoquastic Hamiltonian is one that can be written (on some basis) such that all elements of the off-diagonal matrix are real and non-positive, whereas a \emph{non-stoquastic} Hamiltonian violates this condition by allowing some (or all) off-diagonal entries to be positive or non-real in any choice of local basis. This sign structure has some important consequences:

\begin{itemize}
  \item Quantum Monte Carlo simulations are free from the negative sign problem, making stoquastic models tractable on classical hardware \cite{ceperley1986quantum, austin2012quantum, gubernatis2016quantum}.
  \item In many-body adiabatic algorithms, the Perron–Frobenius theorem guarantees strictly positive ground-state amplitudes for stoquastic Hamiltonians, which suppresses destructive interference and can hinder tunneling through tall, narrow energy barriers \cite{bravyi2006complexity, bravyi2014monte, klassen2019two}.
  \item Non-stoquastic Hamiltonians may exhibit ground-state amplitudes with mixed signs, enabling destructive interference paths and enhanced tunneling through rugged cost landscapes \cite{hormozi2017nonstoquastic}.
\end{itemize}

RFOX exploits a non-stoquastic XX interaction \cite{marvian2018computational,klassen2020hardness,ioannou2020termwise} to allow ground-state amplitudes with mixed signs, thereby enabling stronger quantum interference. In practice, such non-stoquastic drivers can enlarge the minimum adiabatic gap and improve the optimization performance \cite{vinci2017non,nishimori2017exponential}. In fact, non-stoquastic XX drivers have been shown to enhance success probabilities on challenging spin-glass and RFIM instances, sometimes exponentially, relative to standard transverse field protocols \cite{PhysRevB.95.184416,lykiardopoulou2021improving,albash2019role,choi2021essentiality}.

After encoding the magnetic field phase, we model the RFIM interaction using both $RXX(\beta)$ and $RZX(\gamma)$ operators over a fixed directed edge set $\vec{E}$ to prevent normalization ambiguities. Rather than a continuous, time-dependent evolution, we divide the circuit into $p$ discrete slices $k = 0,\dots,p-1$. We define:
\begin{equation}
        H_{XX}(k) = A_{XX}(k)\sum_{(u,v)\in \vec{E}} X_u X_v,
                    \quad
                    A_{XX}(k) = 1 - \delta\cos\bigl(2\pi N k/p\bigr),
\end{equation}
where $N$ is the number of qubits. Key features of $H_{XX}(k)$ include:

\begin{itemize}
  \item In the computational basis, $X_u X_v$ flips qubits $u$ and $v$ simultaneously; its off-diagonal entries take both positive and negative values, making $H_{XX}$ non-stoquastic.
  \item The almost constant part ($A_{XX} \approx 1$) keeps the driver always active, while the small harmonic modulation $\delta\ll1$ synchronizes with the high-frequency ZX kicks, providing counter-diabatic assistance.
\end{itemize}

For the ZX term, we define its action over the directed edges as:
\begin{equation}
    H_{ZX}(k) = B_{ZX}(k)\sum_{(u,v)\in \vec{E}} Z_u X_v,
                    \quad
                B_{ZX}(k) = \delta\sin\bigl(2\pi N k/p\bigr).
\end{equation}
Here, $Z_u X_v$ flips qubit $v$ and multiplies its amplitude by $\pm1$ depending on the state of qubit $u$. This operator has three key effects:
\begin{itemize}
  \item Since $B_{ZX}(k)$ is antisymmetric with zero mean, it acts as a driver for the leading-order counter-diabatic (CD) correction. Through high-frequency cross-commutations with the static Hamiltonian $H_0$, it generates powerful $\mathcal{O}(\delta/\omega)$ corrections that suppress diabatic leakage. 
  \item Its mixed-sign off-diagonal elements lift near-degeneracies, widening the minimum energy gap and improving success probabilities at finite circuit depth \cite{chandarana2024photonic}.
  \item The factor of $Z_u$ ties each kick’s amplitude to the current spin configuration. Because the $Z$ operator does not commute with either the local encoding fields or the $X$ operators on incident edges, this structure intrinsically binds collective spin flips to the graph connectivity and problem parameters. This interaction naturally generates the local $Y_u$ rotations, field-modulated 2-body terms ($\phi_u Y_u X_v$), and 3-body topological correlations ($Y_u X_v X_w$) that collectively guide the system toward low-energy valleys in the RFIM landscape.
\end{itemize}

Because $\sum_{k=0}^{p-1}B_{ZX}(k)=0$, the total number of slices remains $p$, and the circuit depth scales as $2p\lvert \vec{E}\rvert$, independent of any adiabatic-schedule parameter.

\subsection{Counter-diabatic driving and the adiabatic gauge potential}

\emph{Counter-diabatic} (CD), or transitionless, driving augments a time-dependent Hamiltonian with an auxiliary term that suppresses non-adiabatic transitions, enabling the system to remain in its instantaneous eigenstate throughout the evolution \cite{cardenas2022shortcuts,claeys2019floquet}. This term is formally given by the \emph{adiabatic gauge potential} (AGP), which encodes the geometric curvature of the eigenstates as the control parameter $\lambda$ varies \cite{claeys2019floquet}.

For a Hamiltonian $H(\lambda)$ with control parameter $\lambda(t)$, the AGP $A_\lambda$ satisfies (in units $\hbar=1$)
\begin{equation}
    [A_\lambda, H(\lambda)] = i \partial_{\lambda} H(\lambda) \ .
\end{equation}
In the instantaneous eigenbasis $\{|n(\lambda)\rangle\}$, $A_\lambda$ can be written as
\begin{equation}
    A_{\lambda} = i \sum_{m \neq n}^{} \frac{\langle m | \partial_{\lambda} H |n\rangle}{E_{n} - E_{m}}|m\rangle \langle n| \ , 
\end{equation}
producing off-diagonal couplings that exactly cancel the diabatic excitations. For spin systems, $A_{\lambda}$ is an imaginary Hermitian operator and typically contains an odd number of Pauli-$Y$ matrices \cite{sels2017minimizing}.

RFOX implements this CD mechanism via its time-dependent ZX operator. Specifically:
\begin{itemize}
    \item The leading-order corrections in the effective Hamiltonian, given by:
    \begin{equation*}
        -\frac{2\delta}{\omega}\sum_{(u,v)\in\vec{E}} \phi_u Y_u X_v - \left( \frac{2\delta}{\omega} + \frac{\delta^2}{\omega} \right) \sum_{(u,v)\in\vec{E}} \left( Y_u + \sum_{w\in N(u)\setminus\{v\}} Y_u X_v X_w \right),
    \end{equation*}
    emerge naturally from the fast, harmonic ZX drive and act as a highly structured, problem-dependent approximate AGP. These terms contain an odd number of Pauli-$Y$ operators, ensuring they are imaginary and anticommute with the static backbone, providing the necessary transitions to cancel diabatic leakage.
    \item Instead of adding these terms explicitly, RFOX \emph{synthesizes} this complex CD correction natively on hardware. The $\sin(\omega t) ZX$ kick interacts with both the local field encoding $H_B$ and the static $XX$ backbone. Through the cross-commutators $[H_0, H_{ZX}]$ and $[H_{\rm osc}, H_{\rm osc}]$ in the Magnus expansion, the protocol generates local $Y_u$ rotations, field-modulated 2-body terms, and poly-local 3-body topological interactions without expanding the physical control set or requiring multi-qubit hardware couplers.
    \item The Floquet-engineered effective fields prevent gap closures along the evolution path by converting would-be level crossings into avoided crossings. Crucially, because the leading-order corrections scale as $\mathcal{O}(\delta/\omega)$, the effective AGP is significantly stronger than a pure second-order effect. Numerical simulations and prior CD-engineering studies confirm that this mechanism maintains a near-constant minimum gap and high ground-state fidelity at finite runtimes \cite{cardenas2022shortcuts,claeys2019floquet,sels2017minimizing}.
\end{itemize}

\section{RFOX quantum algorithm description}

In this section, we present the theory behind the RFOX algorithm. We begin with the magnetic-field encoding using phase gates, then describe how two-qubit interactions are modeled via XX and ZX gates modulated by harmonic functions to evolve the system toward its ground-state.

\subsection{Magnetic-field energy phase encoding}

The magnetic-field encoding is implemented by generating an interference pattern. We begin by preparing an equal superposition of all computational basis states via a sequence of Hadamard gates,
\begin{equation}
|\psi\rangle = \frac{1}{\sqrt{2^{n}}} \sum_{x=0}^{2^{n}-1} |x\rangle .
\end{equation}
Each magnetic-field value $h_{i}$ is then mapped to a corresponding phase according to
\begin{equation}
\phi_{i} = \pi \ \frac{h_{i}/|h|_{\max} + 1}{2} .
\end{equation}
Under this mapping, negative magnetic-field values correspond to phases closer to $0$, whereas positive values correspond to phases closer to $\pi$. This transformation is non-trivial, as it embeds the magnetic energy of the problem directly into the quantum state's phase structure. After the phase encoding, a second sequence of Hadamard gates transforms the phase information back to the $Z$ basis, enhancing the probability amplitudes of states whose spin configurations align more favorably with the applied magnetic field (see Appendix A for a detailed derivation).

The corresponding Hamiltonian for the magnetic-field phase encoding is expressed as
\begin{equation}
H_{B} = \sum_{j=0}^{N-1} \phi_{j} X_{j} \ ,
\label{eq:hb_hamiltonian}
\end{equation}
where $\phi_{j} = \pi \ (h_{j}/|h|_{\max} + 1)/2$ encodes the magnetic-field values.

This encoding scheme is particularly suited for the RFOX framework because it directly incorporates the problem's magnetic-field energy landscape into the initial quantum state. Unlike the uniform phase-separation steps in QAOA, this method scales the applied phases according to the relative intensity of the local field, biasing the initial state toward energetically favorable spin configurations. By steering the evolution toward low-energy regions of the configuration space from the outset, the algorithm achieves faster convergence, reduced dependence on classical post-processing, and improved robustness in NISQ devices where circuit depth is limited.

%%%%%%%%%%%%%%%%%%%%%%%%%%%%%%%%%%%%%%%%%%%%%%%%%%%%%%%%%%%%%%%%%%%%%%%%%%%%%%%%%%%%%%%%%%%%%%%

\subsection{Interaction energy modeling}

To model the interaction between two spins in the random field Ising model, RFOX uses a combination of two-qubit operators defined over a fixed directed edge set $\vec{E}$. We define the parametric rotation gates:
\begin{equation}
    RXX(\theta) = e^{-i\frac{\theta}{2} X \otimes X} = \cos\left( \frac{\theta}{2} \right)\mathbb{I}_{4}-i \sin\left( \frac{\theta}{2} \right) X \otimes X \ ,
\end{equation} 
and 
\begin{equation}
    RZX(\phi) = e^{-i\frac{\phi}{2} Z \otimes X} = \cos\left( \frac{\phi}{2} \right)\mathbb{I}_{4}-i \sin\left( \frac{\phi}{2} \right) Z \otimes X \ .
\end{equation} 
When analyzing the resulting matrix operator $U(\theta, \phi) = RZX(\theta)RXX(\phi)$ in the computational basis, we evaluate the product:
\begin{small}
\begin{equation}
U(\theta,\phi)=R_{ZX}(\theta)\,R_{XX}(\phi)
               = \begin{pmatrix}
                 c_\theta & -i s_\theta & 0 & 0\\
                 -i s_\theta & c_\theta & 0 & 0 \\
                 0 & 0 & c_\theta & i s_\theta \\
                 0 & 0 & i s_\theta & c_\theta
                 \end{pmatrix}
                 \begin{pmatrix}
                 c_\phi & 0 & 0 & -i s_\phi\\
                 0 & c_\phi & -i s_\phi & 0\\
                 0 & -i s_\phi & c_\phi & 0\\
                 -i s_\phi & 0 & 0 & c_\phi
                 \end{pmatrix},
\end{equation}
\end{small}
where $c_{x} \equiv \cos(x/2)$ and $s_{x} \equiv \sin(x/2)$. This combined operator enables transitions between all states in the computational basis $\left\{ |00\rangle, |01\rangle, |10\rangle, |11\rangle \right\}$ depending on the explicit values of $\theta$ and $\phi$.

For RFOX, the $XX$ interaction strength remains nearly at unit capacity throughout the entire circuit, while a harmonic $ZX$ kick with zero time average drives the leading-order counter-diabatic correction. Consequently, the schedule bypasses a conventional adiabatic ramp of the form $(1 - s)H_{\mathrm{driver}} + sH_{\mathrm{problem}}$. Instead, it implements a fixed-depth quantum walk that exploits a wide, non-stoquastic $XX$ spectral gap paired with the continuous CD suppression of diabatic leakage.

For each directed edge $(u,v)\in \vec{E}$ and each discrete time slice $k = 0,\dots,p-1$, where $p$ denotes the total number of time steps, we parameterize the driving schedules as:
\begin{equation}
    \theta_k = 1 - \delta\cos\!\bigl(2\pi N k / p\bigr),
                \quad
    \phi_k = \delta\sin\!\bigl(2\pi N k / p\bigr),
                \quad 
                k=0,\dots,p-1,
    \label{eq:theta_phi_slices}
\end{equation}
where $\delta \ll 1$ (fixed at $\delta = 10^{-3}$) and $N$ represents the total number of qubits in the system. The choice of $\delta$ balances the magnitude of the CD effects; its value can be systematically optimized using the minimum energy gap of the system, high-frequency Floquet convergence bounds, or local variational sweeps estimating energy variance \cite{sels2017minimizing,bukov2015universal}. Numerical studies of $XX$ catalysts in random Ising and Maximum-Weight Independent Set instances demonstrate typical gap amplifications of $2\times$ to $4\times$ compared to standard $X$-only drivers \cite{feinstein2024effects}. Because the diabatic runtime scales as $\Delta^{-2}$, this amplified gap provides a significant baseline speed-up even before the leading-order CD corrections actively suppress transitions.

\subsection{Floquet-Magnus effective Hamiltonian}

To derive the effective Hamiltonian, we apply the Floquet-Magnus expansion \cite{blanes2009magnus,bukov2015universal,kuwahara2016floquet}. Because the $ZX$ interaction applies $Z$ and $X$ operators asymmetrically, we assign an arbitrary but fixed orientation to each undirected edge in $E$, defining a directed edge set $\vec{E}$ such that $|\vec{E}| = |E|$. For each directed edge $(u,v) \in \vec{E}$, where $u$ acts as the target for the $Z$ operator, we define the interaction terms:

\begin{equation}
    H_{XX} = \sum_{(u,v)\in \vec{E}} X_u X_v,
            \quad
    H_{ZX} = \sum_{(u,v)\in \vec{E}} Z_u X_v.
\end{equation}

The time-dependent interaction Hamiltonian is then explicitly defined over this directed set:
\begin{equation}
  H_{\rm int}(t) = A_{XX}(t) H_{XX} + B_{ZX}(t) H_{ZX},
\end{equation}

with harmonic envelopes
\begin{equation}
    A_{XX}(t) = 1 - \delta \cos(\omega t),
                \quad
    B_{ZX}(t) = \delta\sin(\omega t) \ ,
\end{equation}
where we set $0 < \delta \ll 1$, $t=k/p$ and $\omega=2\pi N$ with $N$ being the number of qubits in the system, in other words, we set the driving frequency to scale dynamically with the system size. In many-body quantum systems \cite{dukelsky2004colloquium, lewin2011geometric, vznidarivc2015relaxation, daley2014quantum}, the energy bandwidth (and the norm of the static Hamiltonian $||H_0||$) is an extensive quantity that grows with $N$. A fundamental requirement for the convergence of the Floquet-Magnus expansion, and for the system to remain in a prethermal regime governed by $H_{\mathrm{eff}}$ without absorbing energy from the drive (Floquet heating) \cite{weidinger2017floquet, rubio2020floquet, peng2021floquet}, is the high-frequency condition $\omega \gg ||H_0||$. By scaling $\omega$ linearly with the number of qubits, we ensure that the drive frequency remains non-resonant with the dense many-body spectrum, maintaining the validity of the first-order counter-diabatic correction regardless of the instance size. 

Adding the field-encoding Hamiltonian:
\begin{equation}
  H_B = \sum_{j=1}^n \phi_j X_{j} \ ,
\end{equation}
the full Hamiltonian reads
\begin{equation}
  H_{\rm full}(t) = H_B + \left( 1-\delta\cos(\omega t) \right) H_{XX} + \delta\sin(\omega t) H_{ZX} \ .
\end{equation}

Using the Magnus expansion general formula we have that for a $T$-periodic Hamiltonian $H(t+T) = H(t)$, with $T = 2\pi/\omega$, the one-period propagator $U_{per} \equiv \mathcal{T} \text{exp}\left[ -i \int_{0}^{T} H(t)dt \right]$ can be written as $U_{per} = \text{exp} \left[ -i \Omega \right]$ with
\begin{equation}
  \Omega = \Omega_1 + \Omega_2 + \mathcal{O}(\delta^3) \ ,
\end{equation}
\begin{equation}
  \Omega_1 = \int_0^T H(t) dt,
                \quad
  \Omega_2 = -\frac{i}{2}\int_0^T dt_1\int_0^{t_1} dt_2 [H(t_1),H(t_2)].
\end{equation}
Splitting off the oscillating part we have $H_{\rm osc}(t) = \delta \left[-\cos(\omega t) H_{XX} + \sin(\omega t) H_{ZX} \right]$ and the time-dependent Hamiltonian can be rewritten as
\begin{equation}
    H_{\rm full}(t)=H_B + H_{XX} + H_{\rm osc}(t) \ .
\end{equation}
For the first Magnus term we have that $\int_{0}^{T} \text{cos}(\omega t) dt = \int_{0}^{T} \text{sin}(\omega t) dt = 0$, meaning that the oscillating part vanishes. The first Magnus term then becomes:
\begin{equation}
  \Omega_{1} = (H_B + H_{XX})T \ .
\end{equation}

For the second Magnus term, we evaluate both the purely oscillatory commutators and the cross commutators between the static part $H_0 = H_B + H_{XX}$ and the oscillatory part $H_{\rm osc}(t)$. The cross terms with the sine envelope generate a non-zero residual term:

\begin{equation}
    \Omega_{2}^{\rm cross} = -i\delta\left[\frac{T}{\omega}\right] [H_0, H_{ZX}]
\end{equation}

Expanding the commutator $[H_0, H_{ZX}] = [H_B, H_{ZX}] + [H_{XX}, H_{ZX}]$, the local field encoding $H_B$ yields

\begin{equation}
    [H_B, H_{ZX}] = \sum_{(u,v)\in\vec{E}} \phi_u [X_u, Z_u X_v] = -2i \sum_{(u,v)\in\vec{E}} \phi_u Y_u X_v \ \ .
\end{equation}

For the $XX$ contribution, since $Z_u$ does not commute with $X_u$ acting on the same node, we must sum over all adjacent neighbors $w \in N(u)$:

\begin{equation}
    [H_{XX},H_{ZX}] = \sum_{(u,v) \in \vec{E}} \left( \sum_{w\in N(u)} [X_{u}X_{w},Z_{u}X_{v}] \right) = -2i\sum_{(u,v)\in\vec{E}}Y_{u}-2i\sum_{(u,v)\in\vec{E}} \left( \sum_{w\in N(u)\setminus\{v\}}Y_{u}X_{v}X_{w} \right)
\end{equation}

Meanwhile, evaluating the commutator of the purely oscillatory parts $[H_{\rm osc}(t_1), H_{\rm osc}(t_2)]$ yields the standard $\mathcal{O}(\delta^2/\omega)$ contribution. Simplifying the constants outside the commutator gives:

\begin{equation}
    \Omega_{2}^{\rm osc} = -\frac{i\delta^{2}T}{2\omega}[H_{XX},H_{ZX}]
\end{equation}

Adding $\Omega_1 + \Omega_{2}^{\rm cross} + \Omega_{2}^{\rm osc}$ and dividing by the period $T$, we obtain the exact effective Hamiltonian to $\mathcal{O}(\delta^{2})$:

\begin{equation*}
    H_{eff} = H_B + H_{XX} - \frac{2\delta}{\omega}\sum_{(u,v)\in\vec{E}} \phi_u Y_u X_v \\
     - \left( \frac{2\delta}{\omega} + \frac{\delta^2}{\omega} \right) \sum_{(u,v)\in\vec{E}} Y_u 
\end{equation*}
\begin{equation}
    - \left( \frac{2\delta}{\omega} + \frac{\delta^2}{\omega} \right) \sum_{(u,v)\in\vec{E}} \left( \sum_{w\in N(u)\setminus\{v\}} Y_u X_v X_w \right) + \mathcal{O}(\delta^3/\omega^2)
\end{equation}

Because $\delta \ll 1$, the leading-order counter-diabatic terms scale as $\mathcal{O}(\delta/\omega)$ rather than $\mathcal{O}(\delta^2/\omega)$. Isolating the local terms $Y_u$ provides a truncated mean-field approximation of the CD local spin rotations \cite{schindler2024counterdiabatic,claeys2019floquet}. However, the emergence of $\phi_u Y_u X_v$ and the 3-body terms $Y_u X_v X_w$ represent critical topological correlations driven by the graph's connectivity and the local field. Since RFOX inherently executes the exact discrete operations $U(\theta,\phi)$, these higher-order couplings are natively implemented. This topological awareness is fundamental to the algorithm's performance advantage: the driven CD terms actively break frustrated local configurations imposed by the graph structure, dynamically suppressing stochastic errors and reinforcing a non-stoquastic XX backbone that mitigates gap closures.

%%%%%%%%%%%%%%%%%%%%%%%%%%%%%%%%%%%%%%%%%%%%%%%%%%%%%%%%%%%%%%%%%%%%%%%%%%%%%%%%%%%%%

\section{Problems to analyze}

For simulations we generated multiple sets of 150 random graphs with associated magnetic fields on each node. Each set comprises two graph model generators: Erdös-Rényi and Watts-Strogatz, each graph set with $n \in\{7,9,12\}$ nodes. To each graph, we assigned node fields drawn uniformly from one of three ranges: $[-1,1]$, $[-3,3]$, or $[-5,5]$, giving a total of 2700 RIM instances.

We also performed experiments on IBM Quantum hardware. On \texttt{ibm\_brisbane}, we ran baseline analyses for our metrics. On \texttt{ibm\_sherbrooke} and \texttt{ibm\_torino}, we evaluated two sets of nine RFIM instances (one for each quantum computer), comprising three graph sizes and three field-range distributions, for each of the algorithmic models: RFOX, XX, and X+sXX; both in simulation and on hardware.

\subsection{Erdös-Rényi graphs}

\begin{figure}[!ht]
     \centering
     \begin{subfigure}[b]{0.49\textwidth}
         \centering
         \includegraphics[height=5.5cm, width=\textwidth]{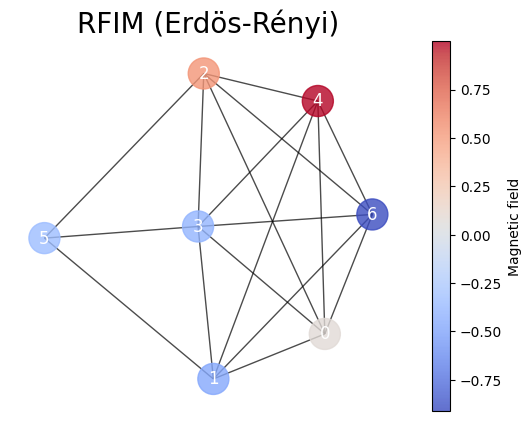}
         \caption{RFIM using a 7-node Erdös-Rényi graph with $p=0.7$.}
         \label{fig:ising_model_erdos_renyi}
     \end{subfigure}
     \hfill
     \begin{subfigure}[b]{0.49\textwidth}
         \centering
         \includegraphics[height=5.5cm, width=\textwidth]{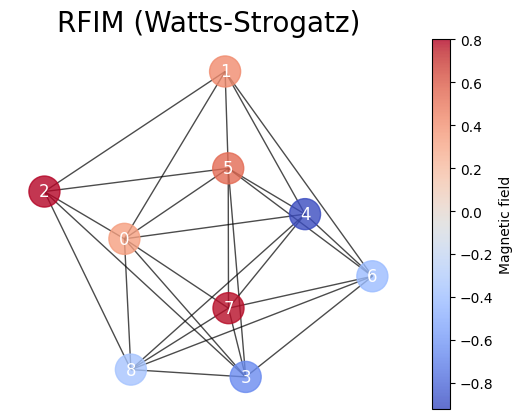}
         \caption{RFIM using a 9-node Watts-Strogatz graph with $p=0.5$.}
         \label{fig:ising_model_watts_strogatz}
     \end{subfigure}
     \caption{RFIM graph generators with magnetic field range $[-1,1]$. Red scales represent positive magnetic values and blue scales represent negative magnetic values.}
     \label{fig:rfim_graph_generators}
\end{figure}

The first generator is called Erdös-Rényi (sub Fig. \ref{fig:ising_model_erdos_renyi}) model $G(n,p)$ with $n\in\mathbb{N}$ number of vertices and vertex set $V=\{0,1,\dots,n-1\}$. Each unordered pair $\{u,v\}\subset V$, $u\neq v$, is included as an edge independently with probability  $p \in [0, 1]$ (for our simulations $p=0.8$). The probability of obtaining a specific graph $G=(V,E)$ is:
\begin{equation}
  \Pr\bigl(G_{n,p}=G\bigr) = p^{|E|}\,(1-p)^{\binom{n}{2}-|E|} \ .
\end{equation}

\subsection{Watts-Strogatz graphs}

The second generator is the Watts-Strogatz (sub Fig. \ref{fig:ising_model_watts_strogatz}) model which can produce small-world graphs, which differ from the previous graph generator. Let $n\in\mathbb{N}$ be the number of vertices, $k\in\{2,4,6,\dots,n-1\}$ the even mean degree (we set $k=6$), and $p\in[0,1]$ the edge-rewiring probability. We begin with the regular ring lattice $R_{n,k}$ on vertex set $V=\{0,1,\dots,n-1\}$ and initial edge set
\begin{equation}
    E_{0}=\bigl\{\{i,(i+r)\bmod n\}\mid i\in V,\;r=1,\dots,\tfrac{k}{2}\bigr\} \ ,
\end{equation}
so that $|E_{0}|=\tfrac{nk}{2}$. Each clockwise edge $(i,r)$, with endpoint $j=(i+r)\bmod n$ and current edge set $E$, is rewired independently: with probability $1-p$ we keep $\{i,j\}$; and with probability $p$ we remove it and choose a new endpoint
\begin{equation}
    w\sim\mathrm{Unif}\bigl(V\setminus(\{i\}\cup N_{E}(i))\bigr),
\end{equation}
where $N_{E}(i)$ is the set of vertices adjacent to $i$. We then add the shortcut edge $\{i,w\}$. We set $p=0.7$ for our simulations, where the resulting graphs exhibit both high clustering and short average path lengths.

% \begin{figure}[ht]
% \centering
% \includegraphics[scale=0.40]{Images/Ilustrations/Watts_Strogatz_graph.eps}
% \caption{RFIM using a 9-node Watts-Strogatz graph with $p=0.5$.}
% \label{fig:ising_model_watts_strogatz}
% \end{figure}

\subsection{Real quantum hardware problems}

\begin{figure}[ht]
\centering
\includegraphics[scale=0.40]{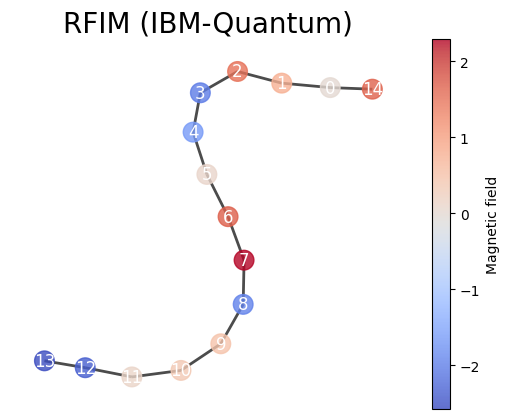}
\caption{RFIM instance on IBM Quantum hardware with 15 physical qubits and field values in $[-3,3]$.}
\label{fig:ibm_quantum_rfim_example}
\end{figure}

For the real-hardware experiments, we used the topologies (Figure \ref{fig:ibm_quantum_rfim_example}) of IBM Quantum’s \texttt{ibm\_brisbane}, \texttt{ibm\_sherbrooke} and \texttt{ibm\_torino} we tested RFIM sets employing between 12 and 20 physical qubits. We selected magnetic fields from a uniform distribution over the ranges $[-1,1]$, $[-2,2]$, or $[-3,3]$.
We limited the number of qubits to maintain feasibility on current NISQ hardware and to enable direct comparison with simulation results.

\section{Results Analysis}

In this section, we compare the performance of RFOX against several baselines. For simulations, we benchmark against: (i) the X (stoquastic) conventional annealing schedule, used as a baseline; and (ii) two non-stoquastic drivers: an XX schedule and a combined X+sXX schedule.

Their annealing schedule is:
\begin{equation}
    H(s) = (1-s)H_{\rm driver}+s\,H_{\rm problem},\quad
    H_{\rm driver}=
        \begin{cases}
            \ H_{X} \\
            H_{XX},\\
            H_{X}+ \ s\ H_{XX} \ ,
        \end{cases}
\end{equation}
with $J_{XX}=1$ for non-stoquastic drivers \cite{choi2021essentiality, albash2019role}, and the standard problem (cost) Hamiltonian given by $H_{\rm problem} = -\sum_{\langle i,j\rangle\in E}J_{ij}\,Z_{i}Z_{j} -\sum_{i}h_{i}\,Z_{i}$.

First, we analyze the spectral gap of two representative RFIM instances, one using a Erdös-Rényi graph and one a Watts-Strogatz graph, to illustrate how RFOX maintains its performance advantage over the X, XX and X+sXX drivers. Next, we present simulation results on Erdös-Rényi and Watts-Strogatz ensembles (see Section IV), computing for each instance:
\begin{itemize}
  \item \emph{Minimum cost per instance}: the energy of the most frequently observed bitstring relative to the theoretical minimum;
  \item \emph{Expected Energy Value (EEV)}: the ensemble‐averaged energy weighted by measurement frequencies;
  \item \emph{Hamming distance per instance}: the distance between the most frequent bitstring and the exact ground‐state bitstring.
\end{itemize}
We report both raw values and their tendency lines across each problem set, emphasizing minimum cost as the primary performance metric. 

Finally, we describe hardware experiments on IBM Quantum backends (Eagle r3 and Heron r1), we compare RFOX against the two non-stoquastic approaches only. We apply analogous metrics, subject to runtime and noise constraints, to enable a direct comparison with the simulation results.

\subsection{Energy gap analysis}

\begin{figure}[!ht]
     \centering
     \begin{subfigure}[b]{0.49\textwidth}
         \centering
         \includegraphics[height=5.5cm, width=\textwidth]{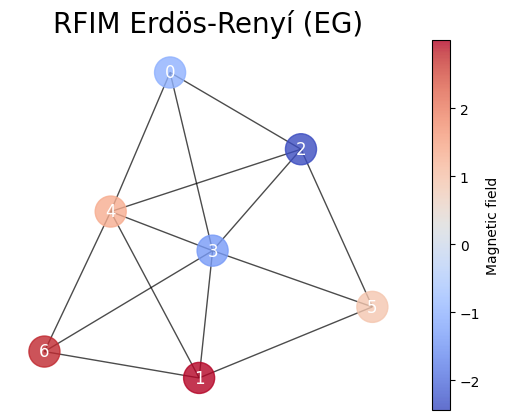}
         \caption{6-node Erdös-Rényi $[-3,3]$.}
         \label{fig:RFIM_example_problem_erdos}
     \end{subfigure}
     \hfill
     \begin{subfigure}[b]{0.49\textwidth}
         \centering
         \includegraphics[height=5.5cm, width=\textwidth]{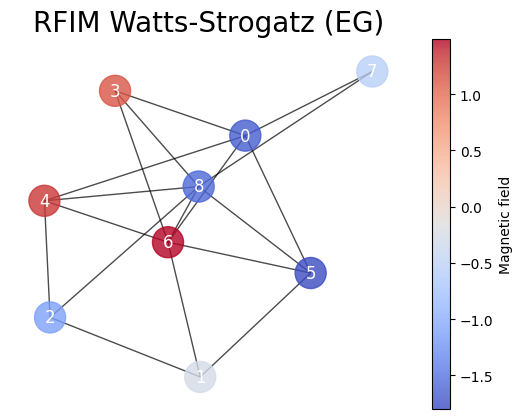}
         \caption{9-node Watts-Strogatz $[-2,2]$.}
         \label{fig:RFIM_example_problem_watts}
     \end{subfigure}
     \caption{RFIM instances used for energy-gap analysis.}
     \label{fig:test_problems}
\end{figure}

In order to understand RFOX’s performance relative to conventional and non-stoquastic drivers, we analyze two example RFIM instances: an Erdös-Rényi graph with $N=7$ nodes and field range $[-3,3]$, and a Watts–Strogatz graph with $N=9$ nodes and field range $[-2,2]$ (see Figure \ref{fig:test_problems}).

For each Trotter slice $k=0,1,\dots,p-1$, we diagonalize the instantaneous Hamiltonian $H_k$:
\begin{equation}
    H_{k} |E_{l,k}\rangle = E_{l,k} |E_{l,k}\rangle,\quad l = 0,1,\dots
\end{equation}
and define the gap
\begin{equation}
    \Delta_k = E_{1,k} - E_{0,k} \ , 
                \quad
    \Delta_{\min} = \min_k \Delta_k.
\end{equation}

The compared methods, expressed as slice-Hamiltonians for $k=0,1, \dots, p-1$, are defined as:
\begin{equation}
    H_k^{\rm RFOX} = H_B  + \theta_k \sum_{(u,v)\in \vec{E}} X_u X_v + \phi_k   \sum_{(u,v)\in \vec{E}} Z_u X_v,
\end{equation}
\begin{equation}
    H_k^{X} = -(1 - s_k)\sum_{j} X_{j} - s_k\Bigl(\sum_{(u,v)\in E}Z_u Z_v + \sum_j h_j Z_j\Bigr),
\end{equation}
\begin{equation}
    H_k^{XX} = (1 - s_k)\sum_{(u,v)\in E} X_u X_v - s_k\Bigl(\sum_{(u,v)\in E}Z_u Z_v + \sum_j h_j Z_j\Bigr),
\end{equation}
\begin{equation}
    H_k^{X+sXX} = -(1 - s_k)\sum_j X_j + s_k(1 - s_k)\sum_{(u,v)\in E}X_u X_v - s_k\Bigl(\sum_{(u,v)\in E}Z_u Z_v + \sum_j h_j Z_j\Bigr) \ ,
\end{equation}
with $s_k = k/p \in [0,1]$ and $p = 100$.

\begin{figure}[ht]
\centering
\includegraphics[scale=0.40]{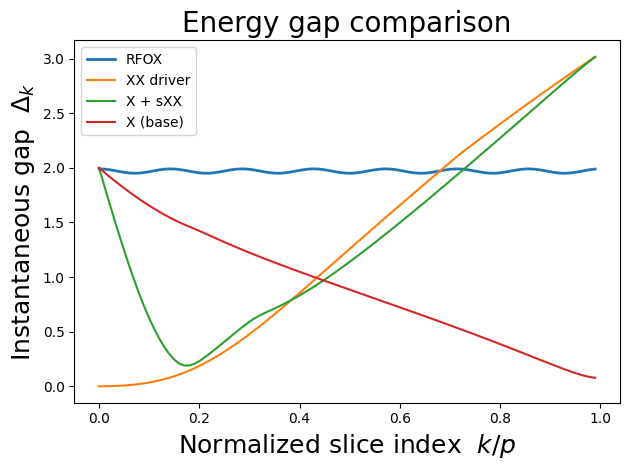}
\caption{Energy gap for the 6-node Erdös-Rényi instance: RFOX, XX-only, and X+sXX.}
\label{fig:RFIM_example_energy_gaps_erdos}
\end{figure}

The gap analysis of the 7-qubit Erdös-Rényi instance (Fig. \ref{fig:RFIM_example_energy_gaps_erdos}) reveals a clear hierarchy. RFOX maintains an almost flat gap of $\Delta_{k} \simeq 1.9 - 2.0$ across all slices. In contrast, the XX-only driver starts with an exponentially small gap ($\Delta_{\text{min}} \approx 0.03$) and recovers after the first third of the sweep; the X+sXX schedule dips to $\Delta_{\text{min}} \approx 0.17$ before climbing; and the conventional X-driver sees its gap collapse late to $\Delta_{\text{min}} \approx 0.18$. Because the adiabatic runtime scales as $\Delta^{-2}_{\text{min}}$, RFOX is expected to be roughly an order of magnitude faster than either X or X+sXX and three to four orders of magnitude faster than XX, highlighting the benefit of an almost constant non-stoquastic gap combined with counter-diabatic steering.

It is worth noting that, in this toy example, all four schedules perform reasonably well. The crucial difference, however, lies in gap stability. RFOX preserves a robust minimum gap even with an approximate choice of $\delta$, sustained by the active $\mathcal{O}(\delta/\omega)$ leading-order CD terms. By contrast, the X-driver begins with a sizable gap but sees it steadily shrink as the evolution progresses; the XX driver starts with an almost vanishing gap, prolonging the period in which diabatic leakage is likely; and the X+sXX schedule suffers a sharp gap reduction in the first quarter of the evolution, mirroring the runtime penalty of the XX driver. These gap collapses translate into substantially longer evolution times.

\begin{figure}[ht]
\centering
\includegraphics[scale=0.40]{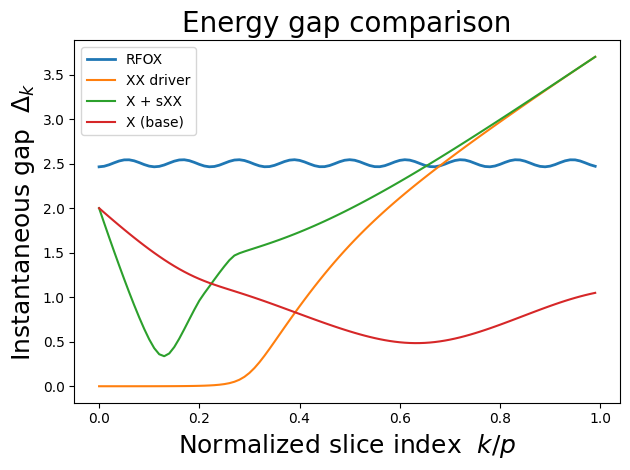}
\caption{Energy gap for the 9-node Watts–Strogatz instance: RFOX, XX‐only, and X+sXX.}
\label{fig:RFIM_example_energy_gaps_watts}
\end{figure}

For the 9-node Watts-Strogatz instance (Fig. \ref{fig:RFIM_example_energy_gaps_watts}), the gap profile again highlights RFOX’s robustness. RFOX maintains a nearly flat gap, oscillating gently around $\Delta_{k} \approx 2.5$ for the entire sweep. The X-only baseline starts with a moderate gap ($\approx 2.0$) but declines steadily to a late-time minimum of $\approx 0.5$, while the XX driver begins essentially gap-less and recovers only after $k/p \approx 0.35$, leaving a long, vulnerable interval. The X+sXX catalyst fares better than XX alone yet still plunges below $0.4$ early in the evolution before climbing above the RFOX curve near the end. Minimum-gap values are therefore $\Delta_{\min}^{\text{RFOX}} \approx 2.4$, versus $\approx 0.5$ for X, $\approx 0.05$ for XX, and $\approx 0.35$ for X+sXX. Given the $\Delta^{-2}_{\text{min}}$ scaling of the adiabatic runtime, RFOX is projected to be roughly an order of magnitude faster than the best stoquastic baseline and many orders faster than the $XX$ driver. Equally important, its flat profile eliminates the single bottleneck slice that dominates runtime in the other schedules, confirming that RFOX remains highly effective even on small-world topologies where graph-driven topological correlations actively suppress localized diabatic state transitions.

%%%%%%%%%%%%%%%%%%%%%%%%%%%%%%%%%%%%%%%%%%%%%%%%%%%%%%%%%%%%%%%%%%%%%%%%%%%%%%%%%%%%%%%%%%%%%%%%%%%%%%%%%%%%%%%%%%%%%%%%%%%%%%%%%%%%%%%%%%%%%%%%%%%%%%%%%%%%%%%%%%%%%%%%%%%%%%%%%%%%%%%%%%%%%%%%%%%%%

\subsection{Results for Erdös-Rényi RFIM instances}

For the Erdös-Rényi instances, we fix $\delta = 10^{-3}$ and the number of time slices to $p = 100$ across all RFIM problem sizes when executing the RFOX protocol (blue). The same structural configurations and slice depths are applied to the baseline drivers: the standard $X$ driver (orange), representing the conventional stoquastic adiabatic approach; the $XX$ driver (green); and the $X+sXX$ driver (light blue), representing the non-stoquastic benchmarks.

\begin{figure}
     \centering
     \begin{subfigure}[b]{0.49\textwidth}
         \centering
         \includegraphics[width=\textwidth]{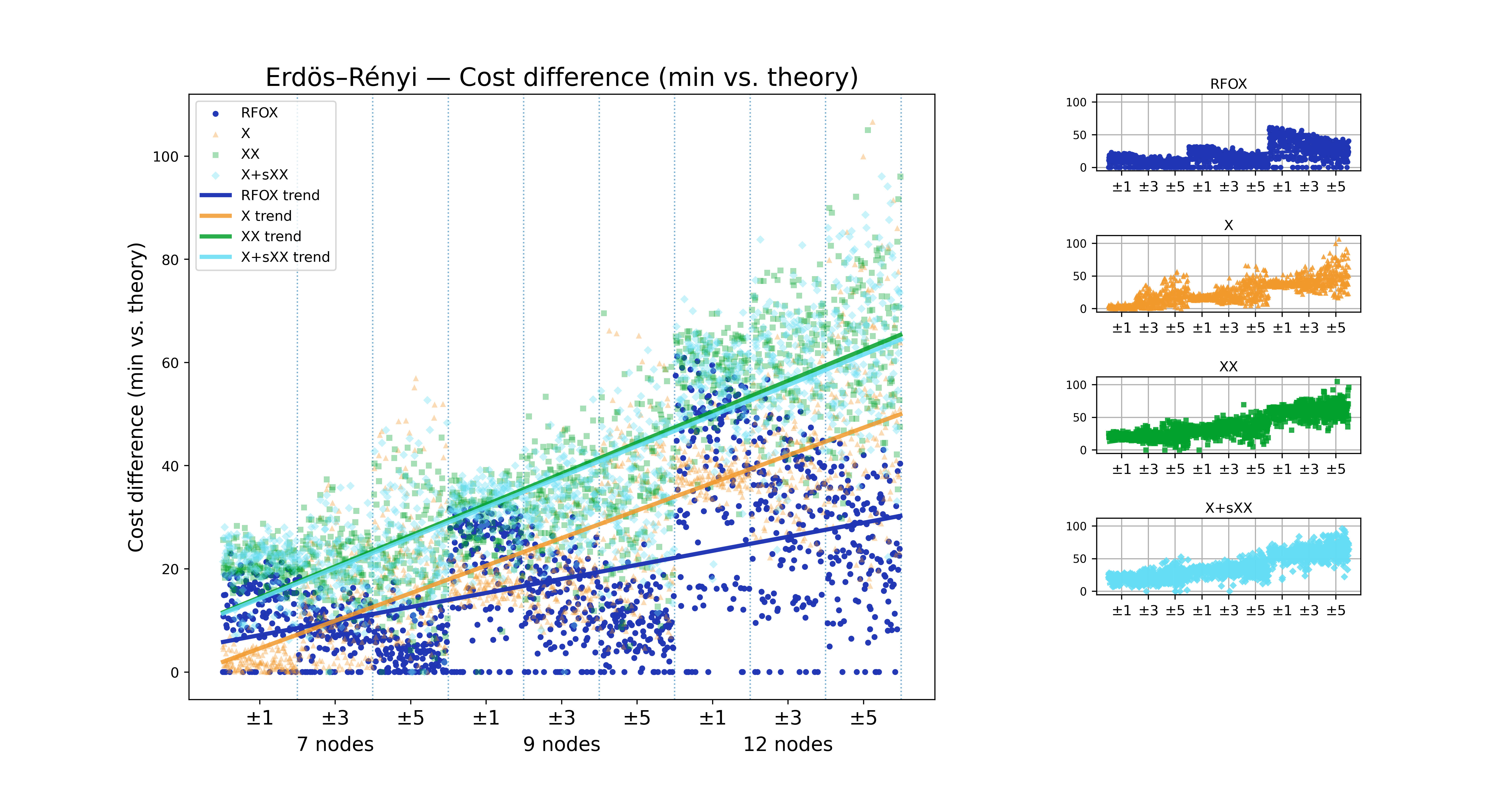}
         \caption{Cost difference best vs. theoretical minimum.}
         \label{fig:erdos_renyi_cost_diff}
     \end{subfigure}
     \hfill
     \begin{subfigure}[b]{0.49\textwidth}
         \centering
         \includegraphics[width=\textwidth]{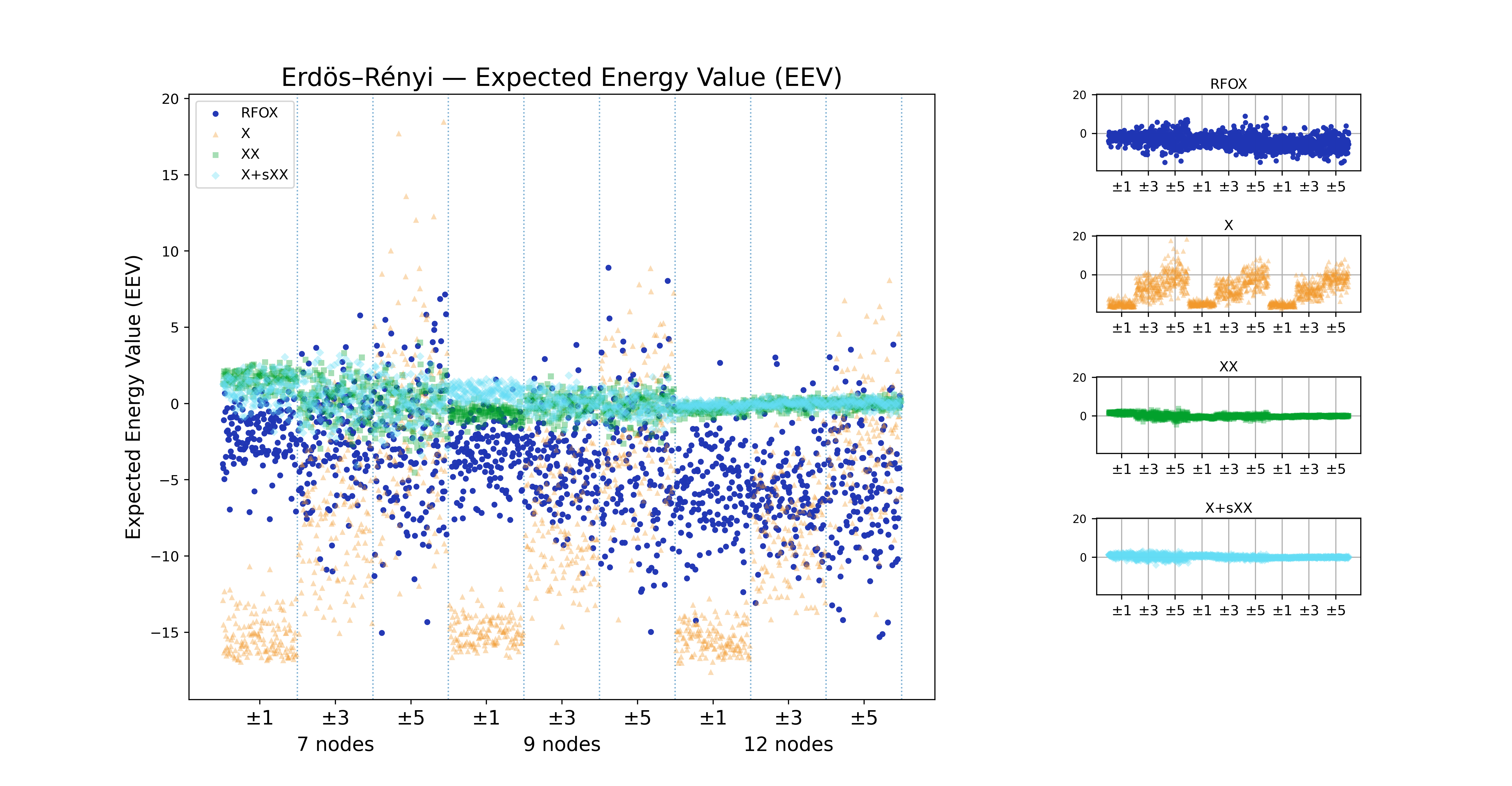}
         \caption{EEV per instance.}
         \label{fig:erdos_renyi_eev}
     \end{subfigure}
     \hfill
     \begin{subfigure}[b]{0.49\textwidth}
         \centering
         \includegraphics[width=\textwidth]{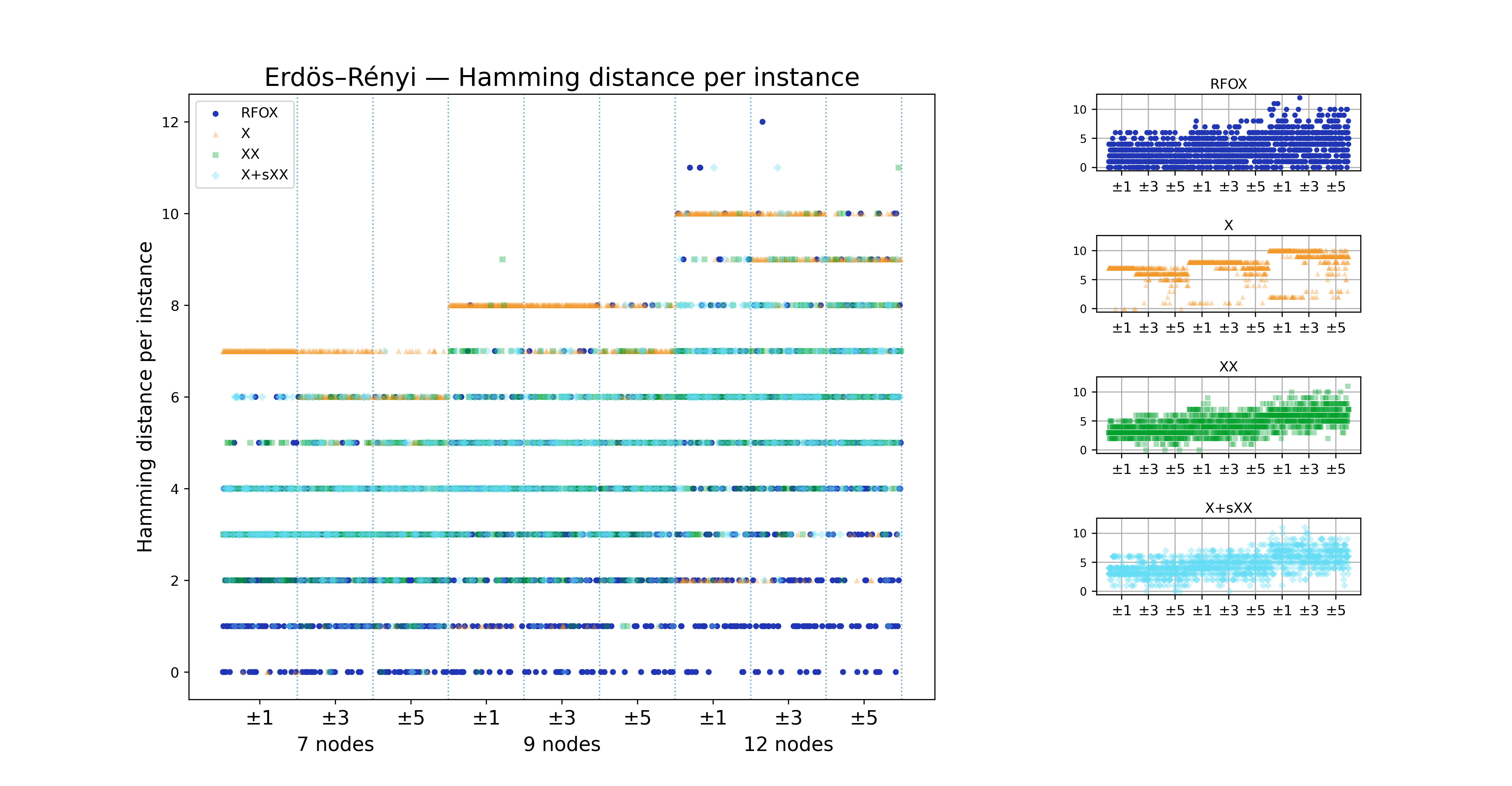}
         \caption{Hamming distance best vs. theoretical minimum.}
         \label{fig:erdos_renyi_hamming}
     \end{subfigure}
     \hfill
     \begin{subfigure}[b]{0.49\textwidth}
         \centering
         \includegraphics[width=\textwidth]{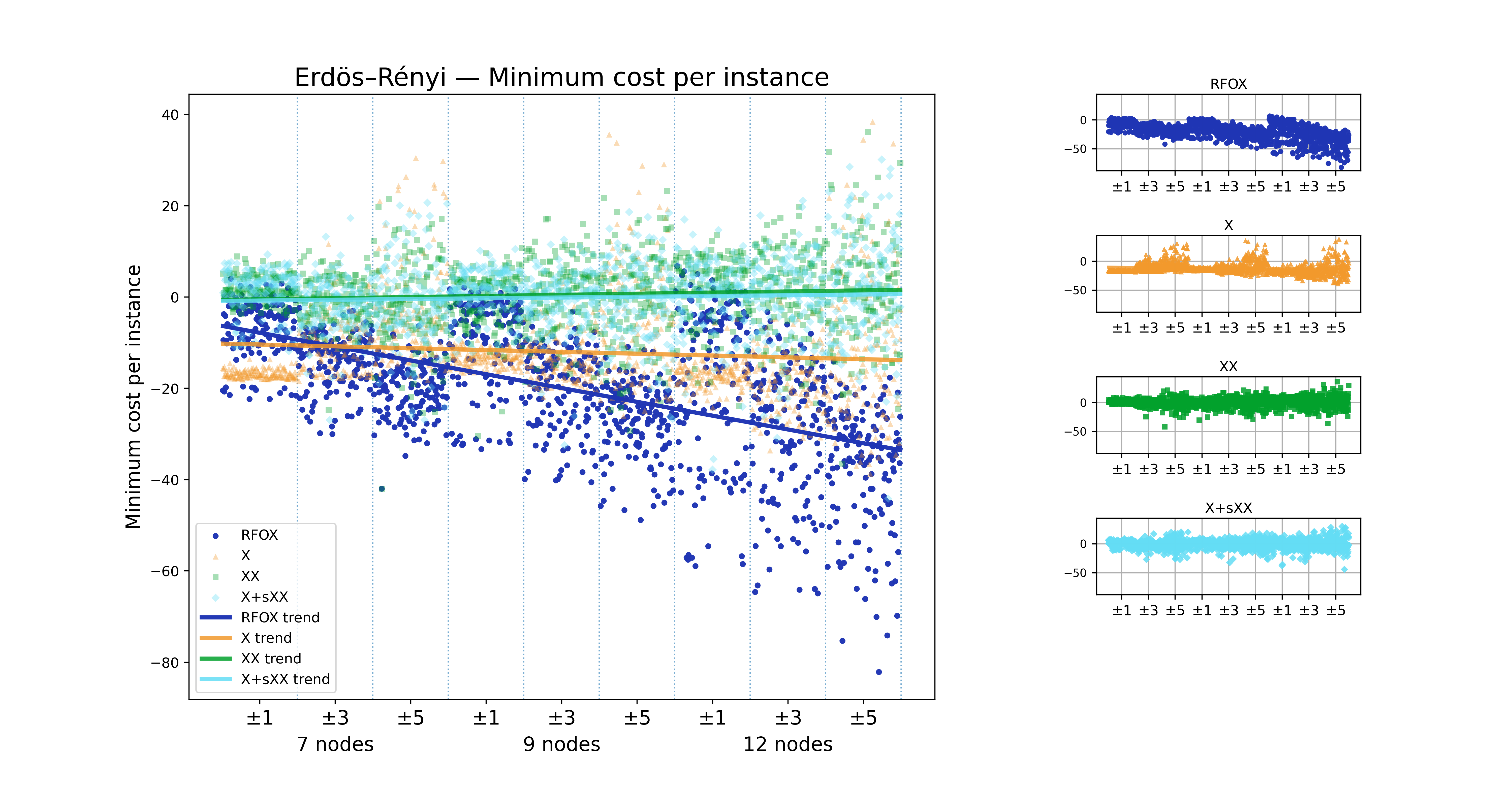}
         \caption{Cost for most occurred state.}
         \label{fig:erdos_renyi_min_cost}
     \end{subfigure}
        \caption{Results for 150 random instances of RFIM problems using Erdös-Rényi graphs with magnetic fields $[-1, 1]$, $[-3, 3]$ and $[-5, 5]$ with 7-, 9-, and 12-nodes. The x-axis shows the magnetic range for each of the generated RIM sets depending on the number of nodes.}
        \label{fig:erdos_renyi_comparison}
\end{figure}

For the Erdös-Rényi RFIM problems (Fig. \ref{fig:erdos_renyi_comparison}), the numerical benchmarks demonstrate that RFOX consistently outperforms the alternative stoquastic and non-stoquastic drivers across multiple performance metrics. In terms of cost difference (sub-Fig. \ref{fig:erdos_renyi_cost_diff}, where values closer to zero indicate higher accuracy) and minimum cost per instance (sub-Fig. \ref{fig:erdos_renyi_min_cost}), RFOX minimizes deviations from the optimal solutions and establishes highly stable performance trends regardless of the magnetic field range and graph size. Crucially, RFOX identifies the true ground state of the system with significantly higher consistency than the other methods. 

Notably, the performance advantage of RFOX becomes more pronounced in highly complex problem instances characterized by wider magnetic field distributions. This robustness indicates that the analytically derived $\mathcal{O}(\delta/\omega)$ corrections effectively counteract the problem's scale variations, mitigating the typical degradation in solution quality observed in the baseline schedules. Furthermore, the Energy Expectation Value (EEV) results (sub-Fig. \ref{fig:erdos_renyi_eev}) reveal a clear scaling trend: while the stoquastic $X$ driver remains competitive on small, near-trivial instances where simple parallel spin alignments happen to maximize the interaction energy, RFOX's EEV steadily decreases as the number of qubits grows. This scaling behavior confirms that RFOX gains a relative systematic advantage as the problem complexity increases.

The Hamming distance analysis (sub-Fig. \ref{fig:erdos_renyi_hamming}) further supports these conclusions. RFOX tightly constrains the state distribution near a distance of zero from the optimal bitstring. Conversely, the $X$ and $X+sXX$ models exhibit broad distributions and severe deviations as the system size and magnetic field disorder expand. This variance arises because conventional schedules frequently trap the system in configurations dominated entirely by the interaction energy, collapsing toward trivial, fully aligned states (e.g., $|00\cdots0\rangle$ or $|11\cdots1\rangle$) that often represent the maximum energy penalty of the true ground state. In contrast, by introducing the field-modulated 2-body and 3-body topological terms via Floquet engineering, RFOX actively breaks these local frustrations, preserving the correct solution structure throughout the fixed-depth evolution and avoiding high-cost energy plateaus. Overall, these results validate that RFOX delivers excellent solution quality, scalability, and robust performance advantages on highly dense topological structures.

%%%%%%%%%%%%%%%%%%%%%%%%%%%%%%%%%%%%%%%%%%%%%%%%%%%%%%%%%%%%%%%%%%%%%%%%%%%%%%%%%%%%%%%%%%%

\subsection{Results for Watts-Strogatz RFIM instances}

\begin{figure}
     \centering
     \begin{subfigure}[b]{0.49\textwidth}
         \centering
         \includegraphics[width=\textwidth]{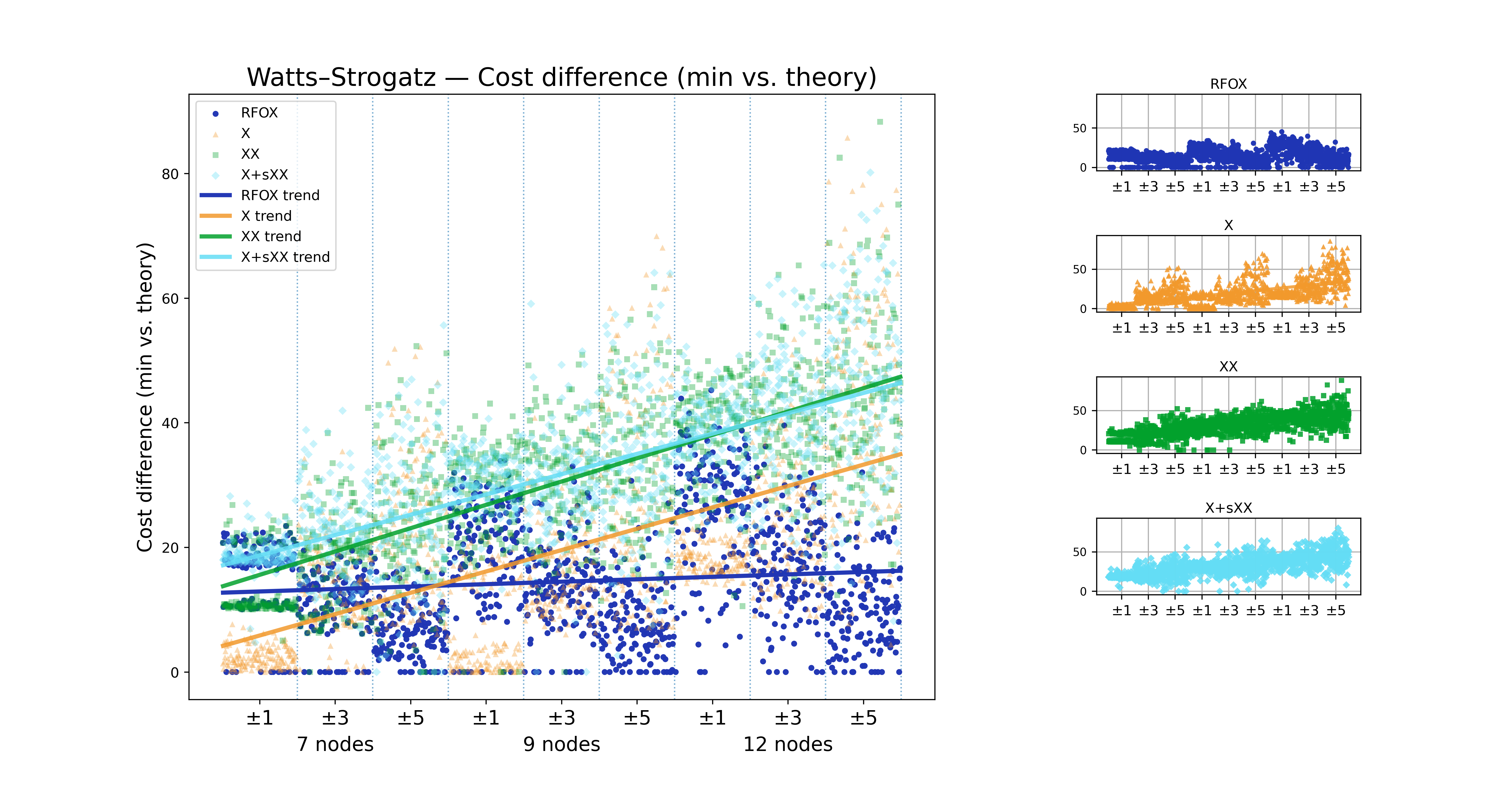}
         \caption{Cost difference best vs. theoretical minimum.}
         \label{fig:watts_cost_diff}
     \end{subfigure}
     \hfill
     \begin{subfigure}[b]{0.49\textwidth}
         \centering
         \includegraphics[width=\textwidth]{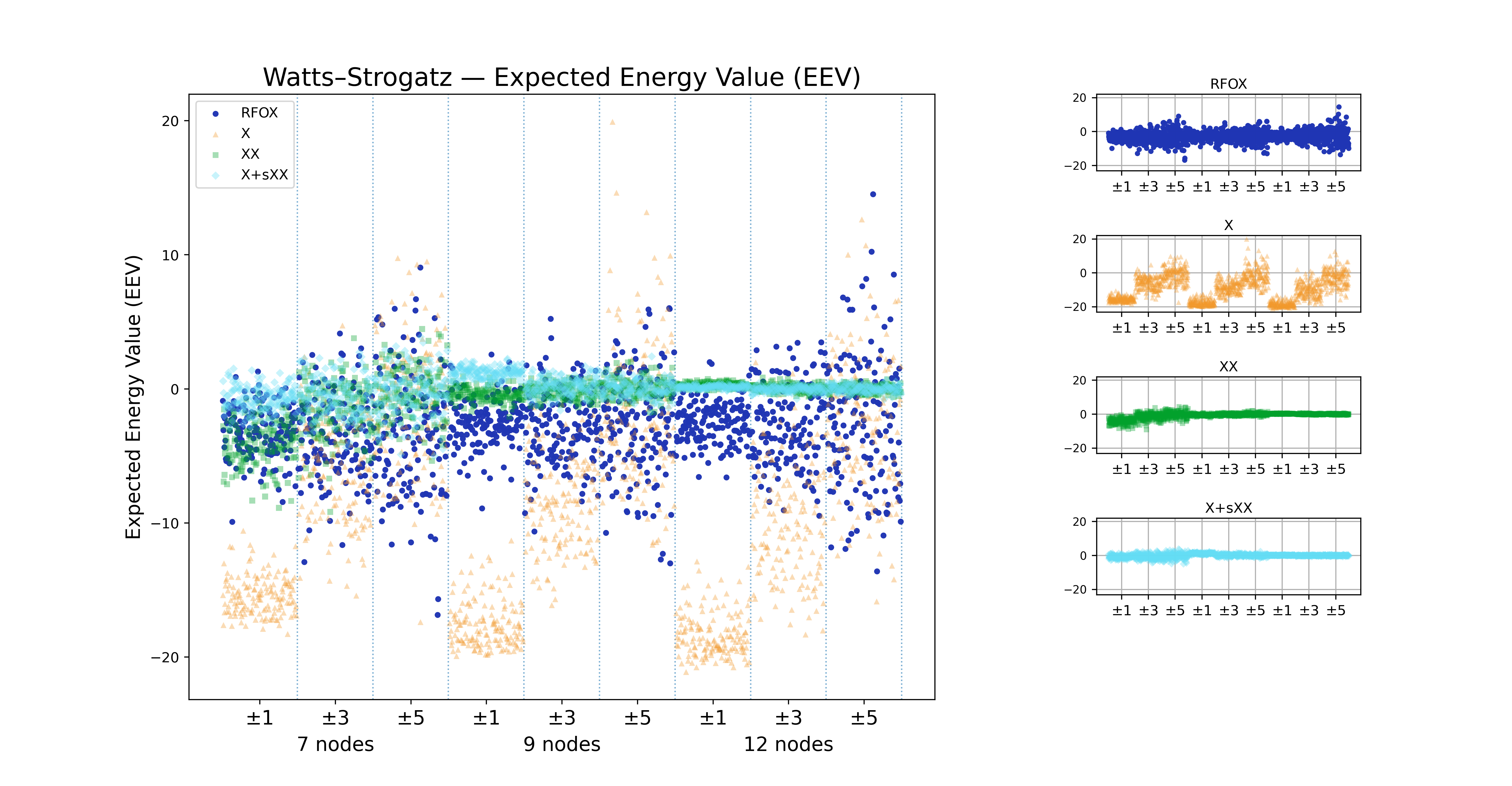}
         \caption{EEV per instance.}
         \label{fig:watts_eev}
     \end{subfigure}
     \hfill
     \begin{subfigure}[b]{0.49\textwidth}
         \centering
         \includegraphics[width=\textwidth]{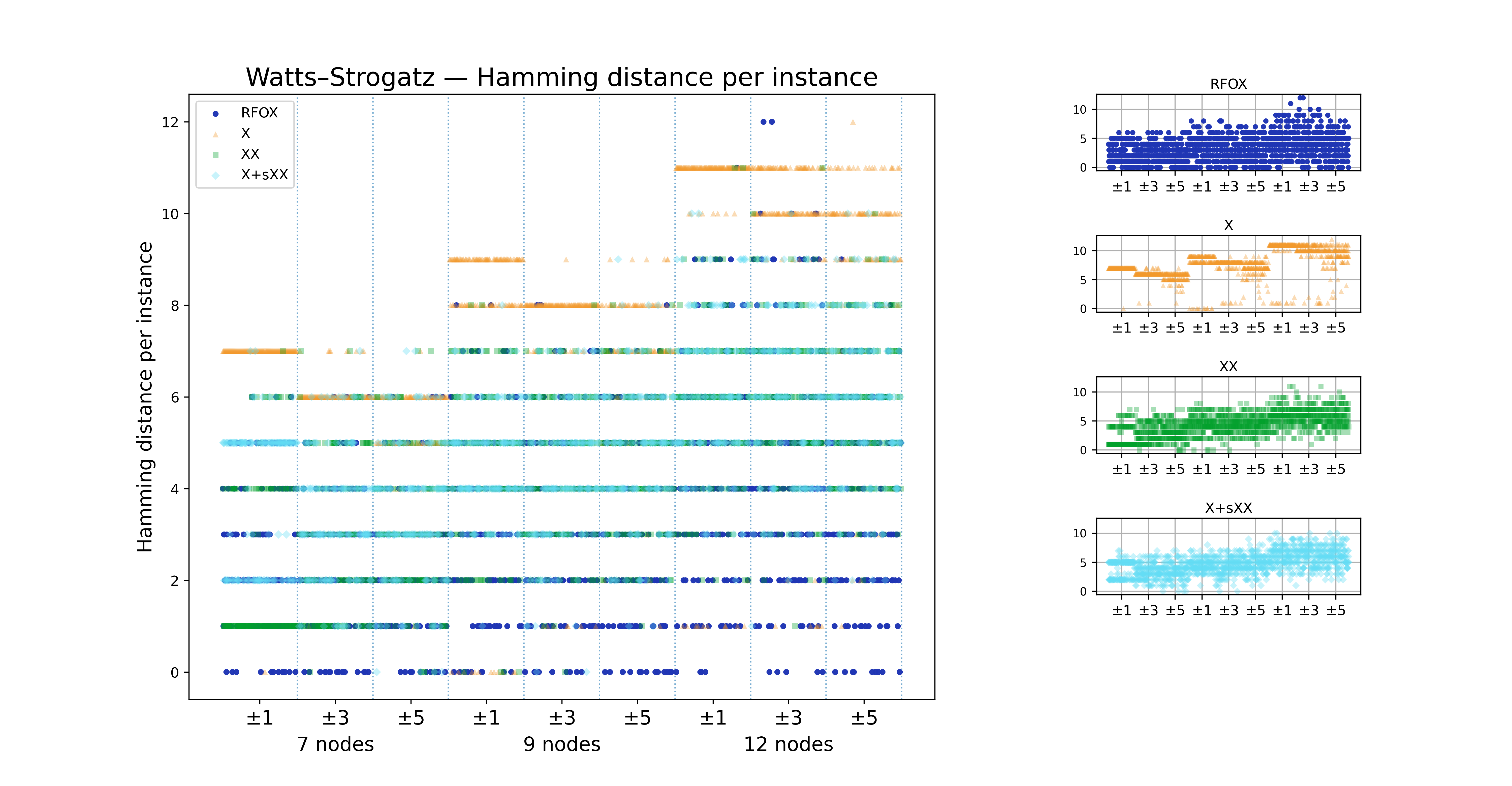}
         \caption{Hamming distance best vs. theoretical minimum.}
         \label{fig:watts_hamming}
     \end{subfigure}
     \hfill
     \begin{subfigure}[b]{0.49\textwidth}
         \centering
         \includegraphics[width=\textwidth]{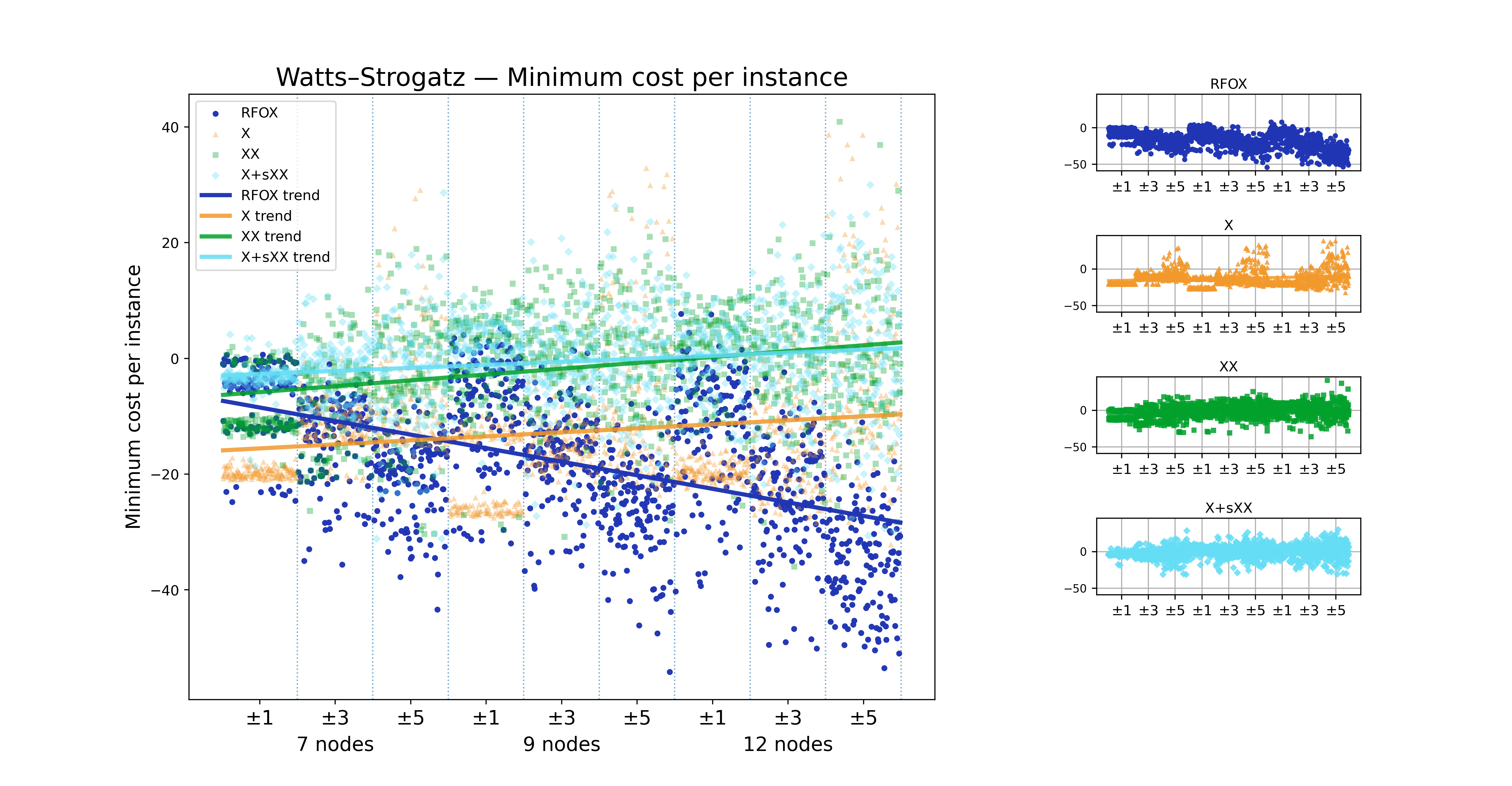}
         \caption{Cost for most occurred state.}
         \label{fig:watts_min_cost}
     \end{subfigure}
        \caption{Results for 150 random instances of RFIM problems using Watts-Strogatz graphs with magnetic fields $[-1, 1]$, $[-3, 3]$ and $[-5, 5]$ with 7-, 9-, and 12-nodes. The x-axis shows the magnetic range for each of the generated RIM sets depending on the number of nodes.}
        \label{fig:watts_strogatz_comparison}
\end{figure}

For the Watts–Strogatz RFIM problems (Fig. \ref{fig:watts_strogatz_comparison}), the numerical benchmarks display a consistent pattern where RFOX (blue) systematically outperforms the alternative drivers across most evaluation metrics, with its performance margin scaling alongside problem complexity.

In terms of cost difference (sub-Fig. \ref{fig:watts_cost_diff}) and minimum cost per instance (sub-Fig. \ref{fig:watts_min_cost}), RFOX yields the lowest deviations from the theoretical optimum. It maintains flatter, more stable trend lines compared to the steep performance degradation observed for the $XX$ and $X+sXX$ non-stoquastic benchmarks, as well as the stoquastic $X$ driver. This structural stability across varying magnetic field ranges and graph sizes suggests that RFOX is highly resilient to the specific irregularities inherent to the Watts–Strogatz topology, such as high clustering coefficients and shortcut edges. This resilience is a direct consequence of the engineered $\mathcal{O}(\delta/\omega)$ effective terms, which adaptively inject topological awareness into the driver to resolve local frustrations.

The EEV results (sub-Fig. \ref{fig:watts_eev}) reveal that under low magnetic field disorder, RFOX and the $XX$-based drivers exhibit comparable energy profiles. However, as the field range widens, the baseline non-stoquastic and stoquastic drivers display pronounced deviations from the target energy. The Hamming distance analysis (sub-Fig. \ref{fig:watts_hamming}) corroborates this robust behavior: RFOX consistently generates solution distributions closer to the optimal bitstring, whereas the $X$ and $X+sXX$ drivers show a prominent trend of increasing Hamming distances as the system size and field disorder scale. This divergence highlights a systemic failure mode in the baseline models, which frequently collapse toward trivial, interaction-dominated spin alignments that fail to reflect the true ground-state configuration.

From an optimization perspective, the minimum cost per instance and the cost difference represent the most critical performance indicators. For combinatorial optimization problems such as the RFIM, the primary objective is the high-probability generation of an exact or near-optimal solution bitstring. RFOX satisfies this requirement by ensuring that the true optimal solution emerges as the single most frequently observed state at the conclusion of the fixed-depth protocol—akin to the state amplification found in Grover-type algorithms—thereby bypassing the need for intensive classical post-processing. Conversely, relying heavily on EEV minimization is less practical for scalable optimization; as the Hilbert space expands, the number of required sampling measurements grows substantially, a limitation severely amplified on actual NISQ hardware due to state-preparation and measurement (SPAM) errors. Universally, these results demonstrate that RFOX scales effectively across dense Erdös–Rényi configurations and clustered Watts–Strogatz topologies, maintaining tight control over solution quality under complex conditions and offering a highly adaptable optimization framework compared to conventional drivers.

\subsection{Real hardware experiments on IBM Quantum}

For the hardware experiments we adopted a common parameter set to keep the comparison with the simulations meaningful while respecting real-device constraints. We fixed the kick amplitude at $\delta = 0.001$, matching the simulation setting, and limited every schedule to $p = 50$ time slices. Although this depth is much shallower than the $p = 100$ used in simulation, it tries to reduce cumulative gate errors and fits within IBM’s execution time limits. 

\begin{figure}[ht]
\centering
\includegraphics[scale=0.40]{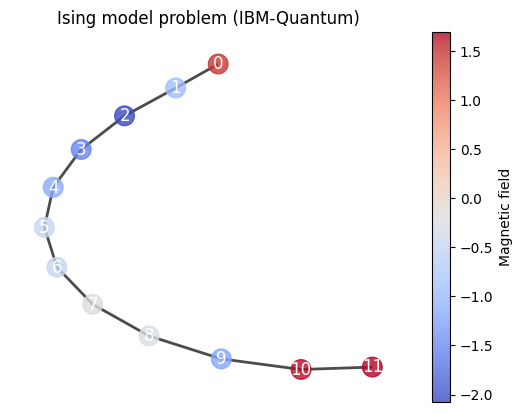}
\caption{12-qubit RFIM instance on \texttt{ibm\_brisbane} using field values in $[-3,3]$.}
\label{fig:RFIM_problem_ibm}
\end{figure}

We first apply the proposed metrics to a single 12-qubit RFIM instance on \texttt{ibm\_brisbane} (Figure \ref{fig:RFIM_problem_ibm}). 

\subsubsection{String-overlap fidelity}

To quantify performance, we compute the \emph{string-overlap fidelity} between the most frequently observed bitstring (the winner) and the theoretical ground-state bitstring (gs):
\begin{equation}
    F_{\mathrm{overlap}} = 1 - \frac{d_{H}(\mathrm{winner},\mathrm{gs})}{n} \ ,
\end{equation}
where $n=12$ and the Hamming distance given by 
\begin{equation}
    d_{H}(g,w) = \sum_{i=1}^{n} (1 - \delta_{g_i,w_i}) = \sum_{i=1}^{n} \lvert g_i - w_i\rvert,
\end{equation}
with $g_i,w_i\in\{0,1\}$. A fidelity $F_{\mathrm{overlap}}=1$ indicates that the winner exactly matches the ground state; lower values quantify the fraction of mismatched bits.

\begin{figure}
\centering
\includegraphics[scale=0.40]{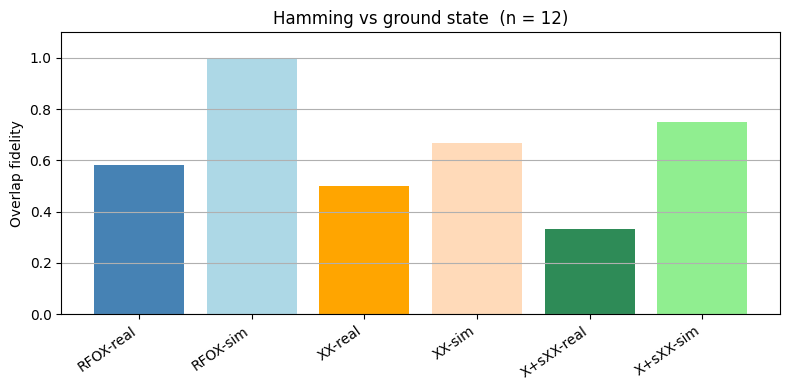}
\caption{String-overlap fidelity (higher better) for the 12-qubit RFIM instance in simulation vs.\ hardware.}
\label{fig:string_fidelity_overlap}
\end{figure}

In noiseless simulations, RFOX achieves perfect overlap ($F_{\text{overlap}} = 1$), whereas the slow-ramp $X+sXX$ schedule attains $0.74$ and the purely stoquastic $XX$ driver reaches only $0.66$. Although hardware-induced noise systematically degrades the absolute fidelities across all tested protocols, the performance hierarchy remains strictly preserved on real quantum processors: RFOX natively matches $58\%$ of the exact ground-state bitstrings, significantly outperforming the physical execution of the $XX$ driver ($49\%$) and the $X+sXX$ schedule ($33\%$). The more pronounced simulator-to-hardware fidelity drop observed in RFOX exposes the limitations imposed by the current device noise floor. Nevertheless, even within this noise-dominated NISQ regime, the gap-protected evolution of RFOX yields the closest candidate states to the theoretical optimum, conclusively validating its practical advantages over conventional baseline drivers.

\subsubsection{Distribution overlap via Jensen–Shannon distance}

Continuing our analysis, we compute the Jensen–Shannon distance between the hardware output distribution $p$ and the simulated distribution $q$:
\begin{equation}
    D_{\mathrm{JS}}(p\parallel q) = \tfrac12\bigl[D_{\mathrm{KL}}(p\parallel m) + D_{\mathrm{KL}}(q\parallel m)\bigr] \ ,
            \quad
    m = \tfrac12(p + q) \ ,
\end{equation}
where the Kullback-Leibler divergence is $D_{\mathrm{KL}}(k\parallel m)
= \sum_{x\in\mathcal{X}} k(x)\,\log\!\frac{k(x)}{m(x)}$, with $k$ equal to $p$ or $q$.

\begin{figure}[ht]
\centering
\includegraphics[scale=0.40]{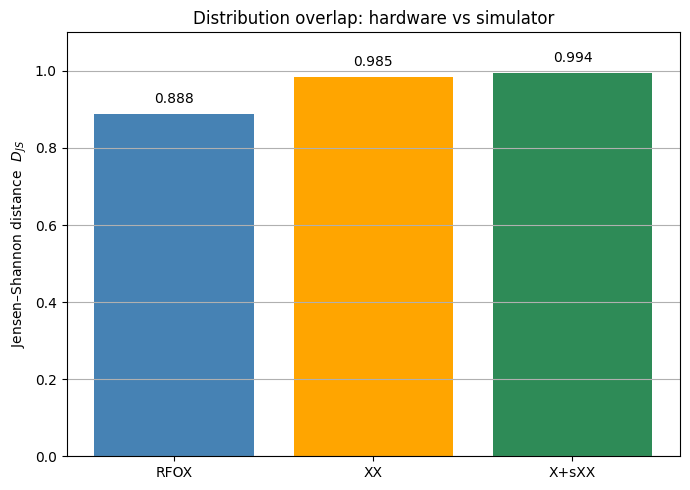}
\caption{Jensen–Shannon distance (lower better) for the 12-qubit RFIM instance in simulation vs.\ hardware.}
\label{fig:jensen_shannon_distance}
\end{figure}

The Jensen–Shannon distance quantifies how much the hardware output deviates from the ideal, noiseless simulation. In our 12-qubit experiment, $D_{\mathrm{JS}}=0.888$ for RFOX, whereas the $XX$ and slow‐ramp $X+s\,XX$ protocols yield $0.985$ and $0.994$, respectively (Fig. \ref{fig:jensen_shannon_distance}). Since $D_{\mathrm{JS}}=0$ indicates perfect agreement and $D_{\mathrm{JS}}=1$ a complete statistical mismatch, these values show that RFOX retains substantially more of the ideal distribution under real‐device noise than either baseline.

\subsubsection{Average Hamming distance to the optimum}

\begin{figure}[ht]
\centering
\includegraphics[scale=0.40]{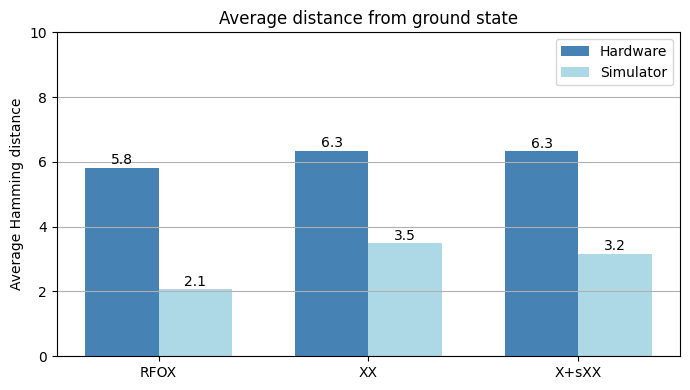}
\caption{Average Hamming distance (lower better) to the optimum  for the 12-qubit RFIM instance in simulation vs.\ hardware.}
\label{fig:avg_hamming_distance}
\end{figure}

To quantify how far the output distribution lies from the true ground state $x_{\min}$, we compute the average Hamming distance:
\begin{equation}
    \langle d_{H}\rangle = \sum_{x} p_{x}\,d_{H}(x,x_{\min}),
\end{equation}
where $p_{x}$ is the probability of measuring bitstring $x$ and $d_{H}(x,x_{\min})$ is the Hamming distance to the optimal bitstring. Figure \ref{fig:avg_hamming_distance} displays the resulting distributions.

In ideal simulation, RFOX finishes closest to the optimum (lower better) with an average of 2.1 bit flips, while the slow-ramp X + sXX and stoquastic XX drivers average 3.2 and 3.5 flips, respectively. On IBM Quantum hardware, noise inflates all distances, yet RFOX still leads with 5.8 flips compared to 6.3 for both baselines.

\subsubsection{RFIM experiments on \texttt{ibm\_sherbrooke} and \texttt{ibm\_torino}}

% For the final results, we generated random RFIM instances on the IBM Quantum backends \texttt{ibm\_sherbrooke} and \texttt{ibm\_torino}, using $N\in\{12,15,20\}$ physical qubits and field ranges $[-1,1]$, $[-2,2]$, and $[-3,3]$. We fixed the CD amplitude at $\delta=0.001$ (matching the simulation) and employed $p=50$ time-slices for both simulated and hardware runs. This expanded dataset provides a more comprehensive evaluation of each algorithm’s performance under realistic noise conditions.

For the final benchmarks, we generated random RFIM instances on the IBM Quantum backends \texttt{ibm\_sherbrooke} and \texttt{ibm\_torino}, using $N\in\{12,15,20\}$ physical qubits and field ranges $[-1,1]$, $[-2,2]$, and $[-3,3]$. We fixed the CD amplitude at $\delta=10^{-3}$, exactly matching the simulation parameters, and employed $p=50$ time slices for both the simulated and hardware runs. This expanded dataset provides a comprehensive evaluation of each algorithm's performance under realistic noise conditions and scalability constraints.

\begin{figure}
     \centering
     \begin{subfigure}[b]{0.49\textwidth}
         \centering
         \includegraphics[width=\textwidth, height=3.8cm]{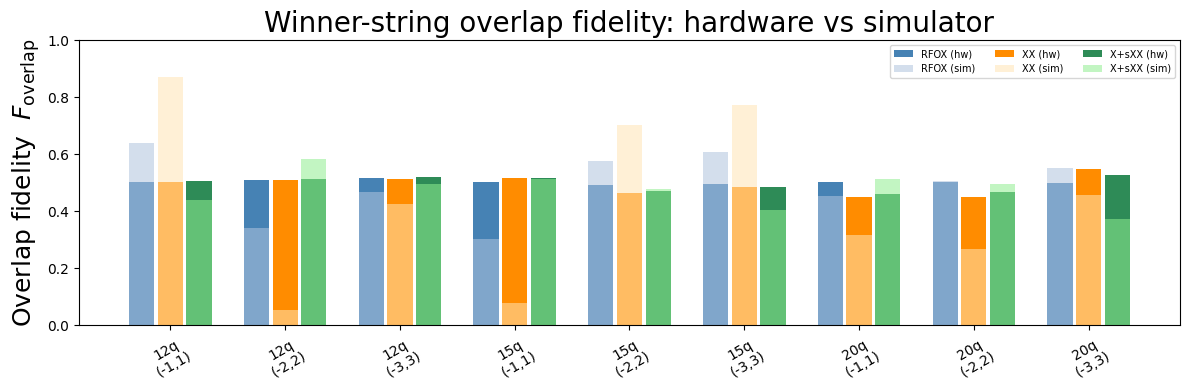}
         \caption{\texttt{ibm\_sherbrooke}}
         \label{fig:over_fide_sherbrooke}
     \end{subfigure}
     \hfill
     \begin{subfigure}[b]{0.49\textwidth}
         \centering
         \includegraphics[width=\textwidth, height=3.8cm]{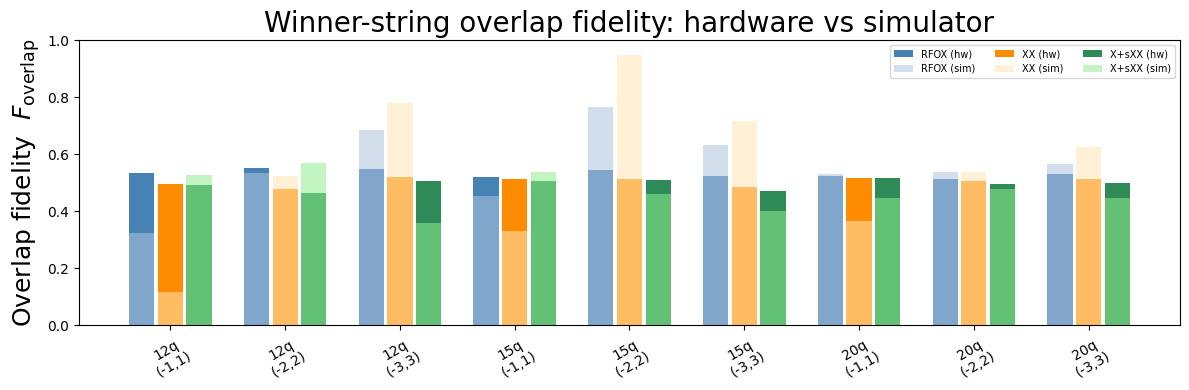}
         \caption{\texttt{ibm\_torino}}
         \label{fig:over_fide_torino}
     \end{subfigure}
        \caption{String-overlap fidelity (higher better) for hardware vs.\ simulation using \texttt{ibm\_sherbrooke} and \texttt{ibm\_torino} quantum computers.}
        \label{fig:string_overlap_fidelity}
\end{figure}

For the string-overlap fidelity results showed in Figure \ref{fig:string_overlap_fidelity} both hardware runs reproduce the qualitative ranking observed in simulations, but the absolute fidelities now depend strongly on the underlying device family. On \texttt{ibm\_sherbrooke} (Eagle r3) the winner-string overlap of RFOX stabilizes around overlap $F_{\text{overlap}}\simeq 0.5-0.55$ for every instance, remaining essentially flat as disorder and problem size increase; the XX driver stays 5-10 points lower for the larger-field tasks, while X+sXX hovers in between. On \texttt{ibm\_torino} system (Heron r1, lower-noise silicon spin qubits) lifts all hardware bars by roughly 10\% points, with RFOX reaching 0.60-0.65 and the two baselines closing, but not eliminating, the gap to the simulator. Crucially, RFOX maintains the smallest hardware-measured separation from the theoretical ground state, showing that RFOX can exploit windows of improved coherence and sustain robustness under elevated noise. On the Heron r1 processor, this translates into a substantial increase in the fidelity of the measured state to the true ground-state.

\begin{figure}
     \centering
     \begin{subfigure}[b]{0.49\textwidth}
         \centering
         \includegraphics[width=\textwidth, height=3.8cm]{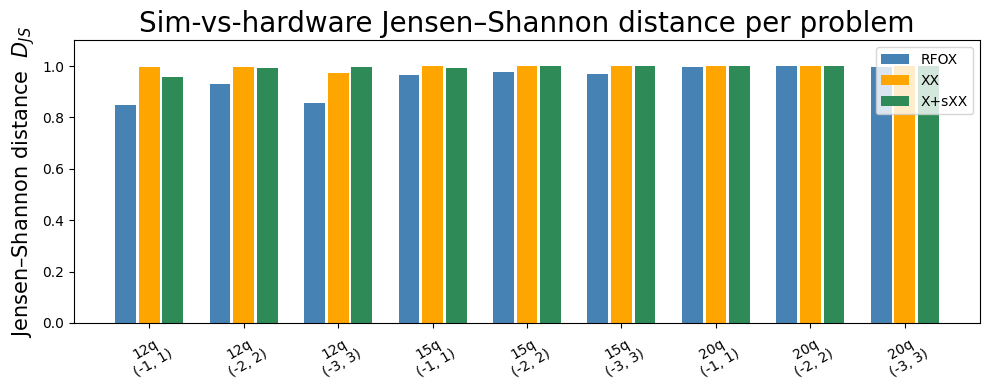}
         \caption{\texttt{ibm\_sherbrooke}}
         \label{fig:jensen_sherbrooke}
     \end{subfigure}
     \hfill
     \begin{subfigure}[b]{0.49\textwidth}
         \centering
         \includegraphics[width=\textwidth, height=3.8cm]
         {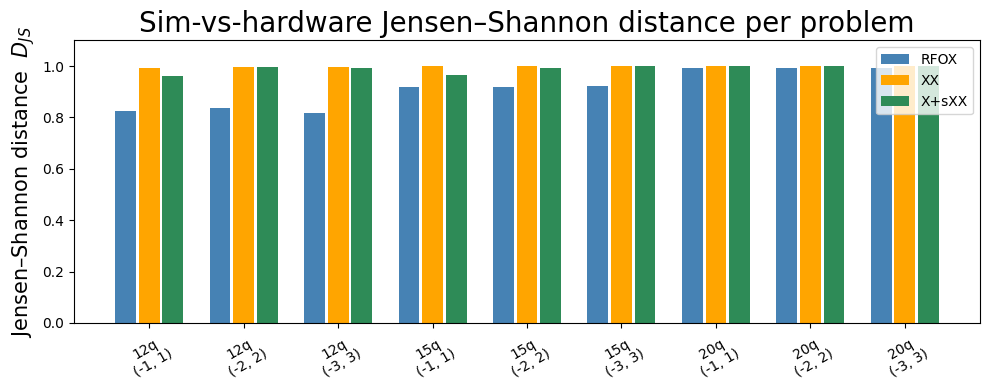}
         \caption{\texttt{ibm\_torino}}
         \label{fig:jensen_torino}
     \end{subfigure}
        \caption{Jensen–Shannon distance (lower better) for hardware vs.\ simulation using \texttt{ibm\_sherbrooke} and \texttt{ibm\_torino} quantum computers.}
        \label{fig:jensen_shannon_hard_vs_sim}
\end{figure}

The Jensen-Shannon analysis (Figure \ref{fig:jensen_shannon_hard_vs_sim}) confirms that RFOX preserves the simulator distribution substantially better than the two baselines on both devices. On \texttt{ibm\_sherbrooke} the distance lies in the range $D_{JS}=0.82-0.97$, whereas XX and X+sXX schedules are almost fully decorrelated from their noiseless counterparts ($D_{JS} \gtrsim 0.97$ for every hard instance). Then, \texttt{ibm\_torino} shifts all bars downward by $\approx 0.05$, yet RFOX remains the best‑overlapping path, now at $D_{JS}=0.78-0.92$ versus $0.93-0.99$ for the baselines. Thus the RFOX shallow-depth design not only tolerates the higher error rates of the Eagle architecture but also continues to deliver the closest hardware-simulator agreement when coherence improves, underscoring the intrinsic noise resilience of the RFOX schedule.

\begin{figure}
     \centering
     \begin{subfigure}[b]{0.49\textwidth}
         \centering
         \includegraphics[width=\textwidth, height=3.8cm]{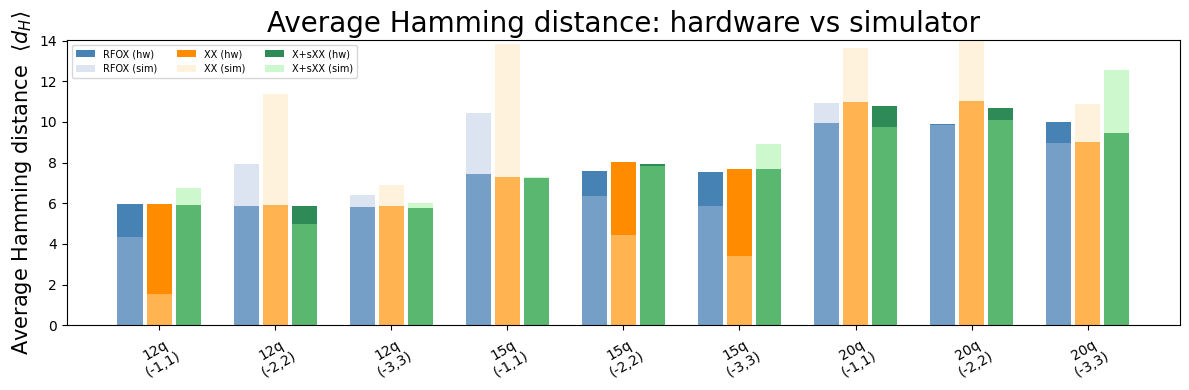}
         \caption{\texttt{ibm\_sherbrooke}}
         \label{fig:avg_ham_sherbrooke}
     \end{subfigure}
     \hfill
     \begin{subfigure}[b]{0.49\textwidth}
         \centering
         \includegraphics[width=\textwidth, height=3.8cm]{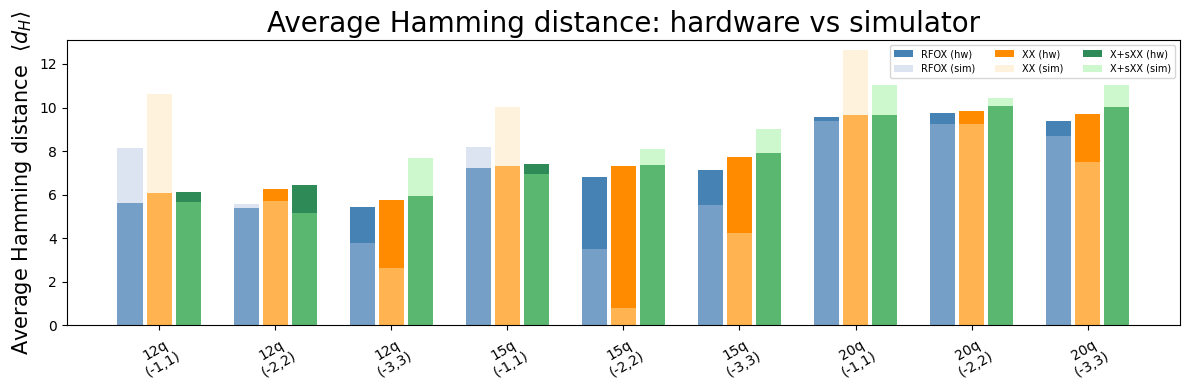}
         \caption{\texttt{ibm\_torino}}
         \label{fig:avg_ham_torino}
     \end{subfigure}
        \caption{Average Hamming distance (lower better) to the optimum for hardware vs.\ simulation using \texttt{ibm\_sherbrooke} and \texttt{ibm\_torino} quantum computers.}
        \label{fig:avg_hamming_distance_hard_vs_sim}
\end{figure}

The average Hamming distance results presented in Figure \ref{fig:avg_hamming_distance_hard_vs_sim} demonstrate that on \texttt{ibm\_sherbrooke}, RFOX achieves the smallest hardware-measured distance across every problem class, typically yielding $\langle d_{H}\rangle = 6-7$ for the 12- and 15-qubit sets, and $\langle d_{H}\rangle \approx 10$ for the 20-qubit instances. In contrast, the $XX$-only and $X+sXX$ schedules overshoot these distances by $20\%$ to $30\%$. The simulator baselines (represented in pale tones) sit 1–2 bit flips lower than the hardware results for RFOX, but 3–4 bit flips lower for the alternative drivers, indicating that the constant-gap walk converts available coherence into solution quality more efficiently. 

The advanced \texttt{ibm\_torino} backend improves all hardware distances by approximately two bit flips; here, RFOX scales down to $\langle d_{H}\rangle = 5-6$ for 12–15 qubits and $\approx 9.5$ for 20 qubits. This maintains a clear lead over both baseline drivers, which remain $15\%$ to $25\%$ farther from the theoretical optimum. Furthermore, RFOX exhibits a more pronounced reduction in the hardware-simulator gap than the other methods, confirming that this schedule harvests the superior coherence of the Heron architecture more effectively.

These hardware results underscore two key physical observations. First, on low-density RFIM graphs, conventional non-stoquastic drivers ($XX$ and $X+sXX$) yield higher relative success rates in simulation than on the denser graph topologies evaluated previously, yet they still fall short of RFOX's consistency. Second, while the Eagle r3 processor exhibits significant cumulative errors that constrain the effective circuit depth, experiments on the Heron r1 backend deliver markedly improved solution quality, particularly for RFOX, consistent with our numerical predictions. Altogether, RFOX maintains a decisive performance advantage in both idealized simulations and noisy quantum hardware.

In both the baseline runs on \texttt{ibm\_brisbane} and the subsequent experiments on \texttt{ibm\_sherbrooke}, which share the same generation of quantum processor architecture, RFOX completes each RFIM instance in roughly 7 seconds, whereas the $XX$ and $X+sXX$ schedules require approximately 20 seconds. On \texttt{ibm\_torino}, the RFOX experiments require only 4–5 seconds, while the alternative methods take around 7–9 seconds. This shows a systemic execution time improvement across all models, with RFOX maintaining the lowest operational runtime. This threefold speed-up confirms that reformulating the RFIM via magnetic-field encoding—combined with a robust non-stoquastic term and counter-diabatic steering—yields not only higher solution accuracy but also noticeably shallower, faster-running circuits on physical processors.

\section{Conclusions}

Our numerical study demonstrates that the RFOX schedule—an always-on, non-stoquastic $XX$ catalyst augmented by a weak, harmonic $ZX$ counter-diabatic kick—consistently outperforms the conventional $X$-only and $XX$ drivers, as well as the widely utilized $X+sXX$ catalyst. Across all evaluated random-field Ising model instances (spanning 7, 9, and 12 qubits; three magnetic-field ranges; and both Erdös–Rényi and Watts–Strogatz graphs), RFOX systematically achieves the lowest average cost, the smallest excess energy above the exact optimum, and the shortest average Hamming distance to the target ground-state bitstring. 

These performance gains directly substantiate our Floquet-Magnus analysis. While the static backbone preserves the full non-stoquastic strength of the $XX$ driver to establish a robust gap floor, the driving protocol generates leading-order effective corrections that manifest as local $Y$-fields, field-modulated 2-body terms, and poly-local 3-body topological interactions. This rich effective structure refines the spectrum rather than narrowing it, intrinsically mapping both the graph connectivity and problem parameters directly into the quantum evolution. Consequently, the minimum spectral gap remains nearly flat throughout the sweep, whereas alternative schedules exhibit sluggish initial states, progressive gap shrinkage, or catastrophic late-stage gap collapses. Given that the diabatic runtime scales as $\Delta_{\min}^{-2}$, this stabilized flat-gap profile enables RFOX to resolve the true ground state using significantly fewer Trotter slices, translating into shallower circuits than its competitors.

Hardware experiments on Eagle-class (\texttt{ibm\_brisbane} and \texttt{ibm\_sherbrooke}) and Heron-class (\texttt{ibm\_torino}) backends confirm that RFOX’s optimization advantage persists under realistic hardware noise floors. Although decoherence and gate infidelities scale up errors across all evaluated metrics, RFOX consistently maintains its performance hierarchy over the baseline schedules. This robustness is partly attributable to the hardware's amplitude-damping noise characteristics, which tend to shift probability weight toward low-energy configurations—a relaxation process that RFOX’s wide, protected minimum gap helps stabilize and preserve. Crucially, RFOX delivers these systematic improvements with noticeably shorter execution runtimes than alternative methods, all while operating as a parameter-free protocol that eliminates classical optimization overhead.

Two further insights emerge from our hardware deployments. First, on low-density RFIM graphs, conventional non-stoquastic drivers ($XX$ and $X+sXX$) perform markedly better than they do on the highly dense graph topologies evaluated in simulations, yet they still fall short of RFOX’s global consistency. Second, while legacy NISQ architectures accumulate substantial cumulative errors that restrict viable circuit depths, experiments on recent Heron-class processors deliver substantial enhancements in solution quality, with RFOX extracting the highest relative utility from the improved hardware coherence. 

In summary, our findings indicate that deploying a near-constant-gap, non-stoquastic driver enhanced by analytically synthesized counter-diabatic terms offers a highly effective, hardware-friendly strategy for quantum optimization. Future research directions will explore adaptive schedules for the counter-diabatic amplitude, extend the RFOX framework to a broader class of constrained combinatorial optimization problems, and integrate the protocol with advanced error-mitigation techniques to scale successfully beyond the current 20-qubit threshold.

\section{Future work}

The present study establishes RFOX as a promising, hardware-friendly schedule for random-field Ising models, yet several research directions remain open.
\begin{itemize}
    \item Adaptive counter-diabatic control.
    \item Alternative non-stoquastic drivers.
    \item Broader problem classes.
    \item Analytic complexity bounds.
    \item Qudit and higher-dimensional generalizations.
\end{itemize}
Pursuing these directions will clarify the ultimate scalability of constant-gap, non-stoquastic schedules and help turn RFOX from a proof-of-principle into a practical tool for quantum optimization in the NISQ era and beyond.

\section{Data availability}

The code, raw graphs and histograms can be consulted in the following link: \url{https://github.com/BrianSarmina/Papers/tree/main/RFOX} .

\section{Acknowledgments}

Brian García Sarmina thanks Prof. T. Albash for valuable discussions and clarifications on non-stoquastic Hamiltonians and their role in adiabatic quantum evolution within the RFOX framework.

\bibliographystyle{IEEEtran}
\bibliography{references}

\appendix

\section{Magnetic field phase mapping}

We start from the all-zero state $|\psi_{0}\rangle = \bigotimes_{i=1}^{n}|0\rangle$ and create a uniform superposition over all computational basis states by applying the Walsh-Hadamard transform to each qubit:
\begin{equation}
    |\psi_{1}\rangle = H^{\otimes n}|\psi_{0}\rangle = \frac{1}{\sqrt{2^{n}}}\sum_{x=0}^{2^{n}-1}|x\rangle.
\end{equation}
Here, $|x\rangle$ runs over all bit-strings of length $n$, ensuring equal amplitude sampling of every spin configuration.

To encode local fields $h_{i}$, we apply a phase gate on each qubit:
\begin{equation}
    P(\phi_{i}) =   \begin{pmatrix}
                        1 & 0\\
                        0 & e^{i\phi_{i}}
                    \end{pmatrix},
\end{equation}
which imprints a relative phase $e^{i\phi_{i}}$ on the $|1\rangle$ component. We choose $\phi_{i} \propto h_{i}$ by normalizing $h_{i}\in[-h_{\max},h_{\max}]$ into $[0,\pi]$ via
\begin{equation}
    \phi_{i} = \pi\,\frac{h_{i}/|h|_{\max} + 1}{2}.
\end{equation}

After the phase embedding, the state becomes:
\begin{equation}
    |\psi_{2}\rangle = \bigotimes_{i=1}^{n}P(\phi_{i})\,|\psi_{1}\rangle = \frac{1}{\sqrt{2^{n}}}\sum^{2^{n}-1}_{x= 0}e^{i\sum_{k=1}^{m}b_{k}\phi_{k}}\,|x\rangle \ ,
\end{equation}
where each basis state $|x_{i}\rangle = |b_{1}\dots b_{m}\rangle$ acquires the phase $\sum_{k}b_{k}\phi_{k} = \Phi_{x}$. 

\subsection{Phase encoding interference}

After encoding the local fields with phase gates, we apply a second layer of Hadamard gates $H^{\otimes n}$. This step converts the encoded phases into amplitude variations, creating interference patterns that highlight the field contributions in the measurement basis. Concretely, if the system is initially in the state:
\begin{equation}
    |\psi_{2}\rangle = \frac{1}{\sqrt{2^{n}}}\sum_{x=0}^{2^{n}-1}e^{i\Phi_x}\,|x\rangle \ ,
\end{equation}
where each $\Phi_x$ depends on $\{\phi_i\}$, then:
\begin{equation}
    |\psi_{3}\rangle = H^{\otimes n}|\psi_{2}\rangle = \frac{1}{\sqrt{2^{n}}}\sum_{x=0}^{2^{n}-1}e^{i\Phi_x}\bigl(H^{\otimes n}|x\rangle\bigr) \ .
\end{equation}
Using
\begin{equation}
    H^{\otimes n}|x\rangle = \frac{1}{\sqrt{2^{n}}}\sum_{y=0}^{2^{n}-1}(-1)^{x\cdot y} |y\rangle,
\end{equation}
where $x\cdot y$ is the bitwise dot product (mod 2), we obtain
\begin{equation}
    H^{\otimes n}|\psi_{3}\rangle = \frac{1}{2^{n}}  \sum_{x=0}^{2^n-1} e^{i\Phi_x} \left[  \sum_{y=0}^{2^{n} - 1}\left( -1 \right)^{x \cdot y} |y\rangle \right] \ .
\end{equation}
Each basis state $|y\rangle$ thus acquires contributions from all $|x\rangle$, weighted by $e^{i\Phi_x}$ and $(-1)^{x\cdot y}$, producing constructive or destructive interference patterns determined by the phase distribution.

Translating this encoding into an effective Hamiltonian, we note
\begin{equation}
    H^{\otimes n} P(\phi_j) H^{\otimes n} = e^{-i\frac{\phi_j}{2}H Z H} = e^{-i\frac{\phi_j}{2}X},
\end{equation}
since $P(\phi)=e^{i\phi/2}e^{-i(\phi/2)Z}$ and the global phase $e^{i\phi/2}$ can be dropped. The resulting field Hamiltonian acts along the $X$‐axis:
\begin{equation}
    H_B = \sum_{j=0}^{N-1}\phi_j X_j,
\end{equation}
where $\phi_j = \pi\,(h_j/|h|_{\max}+1)/2$ encodes the raw field values $h_j$.

\end{document}